\documentclass[%
 reprint,
 amsmath,amssymb,
 aps,
]{revtex4-1}

\usepackage{graphicx}
\usepackage{dcolumn}
\usepackage{bm}
\usepackage[dvipsnames]{xcolor}
\usepackage{hyperref}
\hypersetup{colorlinks=true,linkcolor=black,filecolor=black,urlcolor=black,citecolor=black}

\newcommand{\be}{\begin{equation}}
\newcommand{\ee}{\end{equation}}
\newcommand{\bs}{\begin{equation} \begin{split}}
\newcommand{\es}{\end{split} \end{equation}}

\newcommand{\p}{\partial}

\newcommand{\ts}{\tau_s}
\newcommand{\tm}{\tau_m}

\newcommand{\BigO}{\mathcal{O}}
\newcommand{\dlang}{\langle\langle}
\newcommand{\drang}{\rangle\rangle}
\newcommand{\dblang}{\Big\langle\Big\langle}
\newcommand{\dbrang}{\Big\rangle\Big\rangle}
\newcommand{\dBlang}{\Bigg\langle\Bigg\langle}
\newcommand{\dBrang}{\Bigg\rangle\Bigg\rangle}
\newcommand{\var}{{\rm var}}

\begin{document}

\preprint{APS/123-QED}

\title{Applications of information geometry to spiking neural network behavior}

\author{Jacob T. Crosser$^{1,2}$}
\author{Braden A. W. Brinkman$^{2,1}$}%
 \email{Corresponding author: braden.brinkman@stonybrook.edu}
\affiliation{$^1$Department of Applied Mathematics and Statistics, Stony Brook University, Stony Brook, NY, 11794, USA}
\affiliation{$^2$Department of Neurobiology and Behavior, Stony Brook University, Stony Brook, NY, 11794, USA}

\date{\today}

\begin{abstract}
The space of possible behaviors complex biological systems may exhibit is unimaginably vast, and these systems often appear to be stochastic, whether due to variable noisy environmental inputs or intrinsically generated chaos. The brain is a prominent example of a biological system with complex behaviors. The number of possible patterns of spikes emitted by a local brain circuit is combinatorially large, though the brain may not make use of all of them. Understanding which of these possible patterns are actually used by the brain, and how those sets of patterns change as properties of neural circuitry change is a major goal in neuroscience. Recently, tools from information geometry have been used to study embeddings of probabilistic models onto a hierarchy of model manifolds that encode how model behaviors change as a function of their parameters, giving a quantitative notion of “distances” between model behaviors. We apply this method to a network model of excitatory and inhibitory neural populations to understand how the competition between membrane and synaptic response timescales shapes the network’s information geometry. The hyperbolic embedding allows us to identify the statistical parameters to which the model behavior is most sensitive, and demonstrate how the ranking of these coordinates changes with the balance of excitation and inhibition in the network.
\end{abstract}

\maketitle


\section{Introduction}\label{sec:Intro}

A major obstacle to understanding the computational underpinnings of the brain is the high dimensionality of its inputs---environmental stimuli such as light and sound---and its outputs---the activity of neurons and the organismal behaviors they enact \cite{MizusakiODonnel2021}. The behavioral space of a neural circuit with $N$ neurons is unmanageably large: the number of possible spike train patterns such a network can in principle produce over a trial of time length $T$ divided into time bins of size $\Delta t$ is of the order ${\sim}2^{NT/\Delta t}$, assuming at most one spike per time bin. As $\Delta t \rightarrow 0$, this output space becomes infinite-dimensional. However, the behavior of a neural population does not occupy this entire space, as activity is correlated across time and neurons, and the actual behavior of any given neural circuit constitutes just a subset of all possible observations. Perhaps surprisingly, analysis of experimental data has repeatedly found that under many conditions collective neural activity is low dimensional, often comprising less than ${\sim}10^2$ dimensions of this infinite space \cite{MazorLaurent2005,PillowSimoncelli2006,GanguliEtAl2008,CunninghamByron2014,SadtlerEtAl2014,ArcherEtAl2015,MazzucatoFontaniniLaCamera2015,MazzucatoFontaniniLaCamera2016,GaoEtAl2017,MurrayEtAl2017,WarnbergKumar2017,MacdowellBuschman2020}.

Theoretical and computational work in neuroscience has largely focused on investigating the role that synaptic connections between neurons play in shaping the possible activity patterns of a network \cite{FieldChichilnisky2007,TrousdaleEtAl2012,HuEtAl2013,OckerEtAl2017,OckerEtAl2017statistics,BatistagarciaramoFernandezverdecia2018,CurtoMorrison2019}, which can be represented by manifolds (hyper-surfaces) in the behavioral output space of a neural circuit. These manifolds are complicated by the fact that many distinct neural circuits give rise to essentially identical patterns of activity \cite{PrinzBucherMarder2004,MarderBucher2007,MarderGoeritzOtopalik2015,CropperDacksWeiss2016}, meaning many different configurations are mapped to nearby points on these manifolds. 
Understanding how network activity and function changes as network properties or states change is a fundamental problem in neuroscience, and learning how to manipulate this activity most efficiently could lead to new and more effective treatments of neurological disorders.

Taming the possible behavioral repertoires of neural circuits by brute force simulations of network activity is computationally expensive and impractical for circuits larger than a few neurons. The tools of information geometry offer a possible means of representing network activity in an abstract way, but one that is easier to apply to larger networks and begin to understand how to most effectively move a network through its parameter space to achieve desired output behaviors \cite{NakaharaAmari2002,WuAmariNakahara2002,AmariEtAl2003,AmariParkOzeki2006,ShimazakiEtAl2012,AmariKarakidaOizumi2019,AmariKarakidaOizumi2019-2,KarakidaAkahoAmari2020}. Note that this use of information geometry is a means of understanding the structure of complex models themselves, in contrast to applications of information theory in neuroscience as a modeling tool for understanding sensory coding \cite{barlow1961possible,laughlin1981simple,van1992theory,atick1992could,rieke1999spikes,averbeck2006neural,wang2012optimal,moreno2014information,gjorgjieva2014benefits,kastner2015critical,brinkman2016efficient,zylberberg2017robust,pruszynski2019language}.

Many models in complex biology generate a hierarchy of ``hyperribbons'' in their behavioral output space. These hyperribbons are manifolds with a few long directions of the manifold, representing ``stiff'' directions that separate disparate activity states, and many thin directions, which represent ``sloppy'' directions that describe networks with very similar behavior \cite{transtrum2010nonlinear,TranstrumMachtaSethna2011,MachtaChachraTranstrumSethna2013,transtrum2014model,transtrum2015perspective,GutenkunstEtAl2007,QuinnEtAl2019,TeohEtAl2020}. 
These model manifolds come equipped with a natural metric that measures a sense of difference in behavior that is like a distance. We can determine the combinations of parameters that predict the bulk of the behavioral space of the network by identifying these model manifolds. This opens a path for better understanding of how to manipulate network properties to tune a circuit between different regimes of behavior.

In this work, we apply tools of information geometry to models of neural circuitry, and investigate how the balance of single-neuron properties and the properties of the synaptic connections between neurons shape the hierarchy of possible behaviors of the networks. Specifically, we study how changing the membrane and synaptic time constants of the networks shape the manifold hierarchy. We also investigate how adjusting the balance of excitation and inhibition in the network change the rankings of the different hierarchical modes of the behavioral space. Previous work applying ideas from information geometry to neuroscience have primarily to study abstracted representations of spiking networks \cite{NakaharaAmari2002,ShimazakiEtAl2012}, networks of rate models \cite{AmariParkOzeki2006,AmariKarakidaOizumi2019,AmariKarakidaOizumi2019-2,KarakidaAkahoAmari2020}, or neural field and pool models \cite{WuAmariNakahara2002,AmariEtAl2003}. By contrast, the work presented in this paper studies a class of leaky integrate-and-fire neurons---a commonly used modeling framework---with explicit consideration of some biophysical properties of individual neurons to make closer contact with the biological reality of neural systems. This is done by leveraging the specific properties of recently developed tools in information geometry \cite{QuinnEtAl2019,TeohEtAl2020}.
We organize the paper as follows: in Sec.~\ref{sec:Models} we introduce the class of stochastic spiking models we will be working with and the reduction to a population-based formalism. Then, in Sec.~\ref{sec:InfoGeo}, we give a self-contained explanation of the ``isKL'' embedding method introduced by \cite{TeohEtAl2020}, and how it applies to our population model. We detail the results of the application of the isKL method in Sec.~\ref{sec:Results}, and finally discuss the interpretation and significance of our results and methodology in Sec.~\ref{sec:Discussion}. 

\section{Models}\label{sec:Models}

\subsection{Nonlinear Hawkes process}\label{sec:NLHP}

To model the spiking dynamics of individual neurons, we consider a nonlinear Hawkes process \cite{OckerEtAl2017,BrinkmanEtAl2018}
\begin{subequations}
\label{eqn:HawkesProc}
\begin{align}
    \frac{d V_i}{dt} = &-\tm^{-1}(V_i-\varepsilon_i)+I_{i}\nonumber\\
    & +\ts^{-1}\left(\mu_{\rm ext}-J_{\rm  self}\dot{n}_i(t)+\sum_{j=1}^nw_{ij}\dot{n}_{j}(t) \right)\label{subeqn:HawkesMembr}
\end{align}
\begin{equation}
 \dot{n}_i(t)dt\sim {\rm Poiss}[\phi(V_i(t))dt],\label{subeqn:HawkesSpk}
\end{equation}
\end{subequations}
where $V_i$ is the membrane potential of neuron $i$, $\varepsilon_i$ is the leak reversal potential, $w_{ij}$ is the strength of a synaptic connection from neuron $j$ to neuron $i$, and $-J_{\rm self}$ is an inhibitory self-coupling to implement post-spike refractory dynamics. The two currents $\mu_{\rm ext}$ and $I_{i}$ represent an average current received from an external network and an experimentally injected current that differs by neuron, respectively. The process $\dot{n}_i(t)$ is the spike train of neuron $i$, and $\phi(V_i(t))dt$ is the instantaneous firing rate nonlinearity that determines a Poisson event rate conditioned on the membrane potential of a given neuron. For the specific models studied here, $\phi(x) = \frac{1}{2}(x+\sqrt{x^2+1/2})$. Finally, $\tm$ and $\ts$ are modulated parameters corresponding membrane and synaptic timescales, respectively. Eqn.~\ref{subeqn:HawkesMembr} of this model assigns leaky integration dynamics to the membrane potential of each individual neuron, while Eqn.~\ref{subeqn:HawkesSpk} assigns conditionally Poisson spiking dynamics to each neuron. Taken together, this model can be thought of as a soft-threshold leaky integrate-and-fire system.

Foreshadowing the coming analysis, we note that analytically calculating the statistical properties of the models in Eqn.~\ref{eqn:HawkesProc} is generally intractable, and to make headway we will implement a Gaussian-process approximation of the network dynamics around the mean-field activity. 

We can obtain a mean-field approximation of the steady-state solution for the membrane potential dynamics in Eqn.~\ref{subeqn:HawkesMembr} by marginalizing out the spiking dynamics and assuming the distribution is sharply peaked around the most probable path of $V_i(t)$. Assuming the network achieves a steady state at long times, this procedure gives us a set of transcendental equations that can be solved numerically:
\begin{align}
    \footnotesize
    V_i^{\rm mf}&=\varepsilon_I+\tm I_{i}\nonumber\\
    &~~~+\frac{\tm}{\ts}\left(\mu_{\rm ext}-J_{\rm self}\phi(V_i^{\rm mf})+\sum_{j}w_{ij}\phi(V_j^{\rm mf})\right), \normalsize \label{eqn:HP_MembrMF}
\end{align}
where the $V_i^{mf}$, the solutions of these equations, are the mean-field predictions of the steady-state values of membrane potentials, with $\phi(V_i^{mf})$ the corresponding mean-field prediction of the firing rates. We find the solutions to these transcendental equations using a forward-Euler integration scheme.

Following the prescription of Ref.~\cite{ChowBuice2015,Brinkman2023pre}, the time-dependent distribution of model behaviors described in Eqn.~\ref{eqn:HawkesProc} can be written in the form of a path integral,
\begin{equation}
    P[\mathbf{V}(t),\dot{\mathbf{n}}(t)] = \int\mathfrak{D}[\Tilde{\mathbf{V}},\Tilde{\mathbf{n}}]e^{-S[\Tilde{\mathbf{V}},\mathbf{V},\Tilde{\mathbf{n}},\dot{\mathbf{n}}]},
\label{eqn:pathint}
\end{equation}
with an action $S$ given by
\begin{widetext}
    \begin{align}        S[\tilde{\mathbf{V}},\mathbf{V},\tilde{\mathbf{n}},\dot{\mathbf{n}}]=& \int dt\,\,\sum_{i=1}^n\left\{\tilde{V}_i\left [\dot{V}_i+\frac{V_i-\varepsilon_i}{\tm}-I_{i}-\ts^{-1}\left(\mu_{\rm ext}-J_{\rm  self}\dot{n}_i(t)+\sum_{j}w_{ij}\dot{n}_{j}(t) \right) \right]\right.\nonumber\\
        & \hspace{5.0cm} + \tilde{n}_i(t)\dot{n}_i(t)- \left( e^{\tilde{n}_i(t)}-1\right) \phi(V_i)   \Bigg \}
    \label{eqn:full_action}
    \end{align}
\end{widetext}

\begin{figure}[t]
\includegraphics[width=0.8\linewidth]{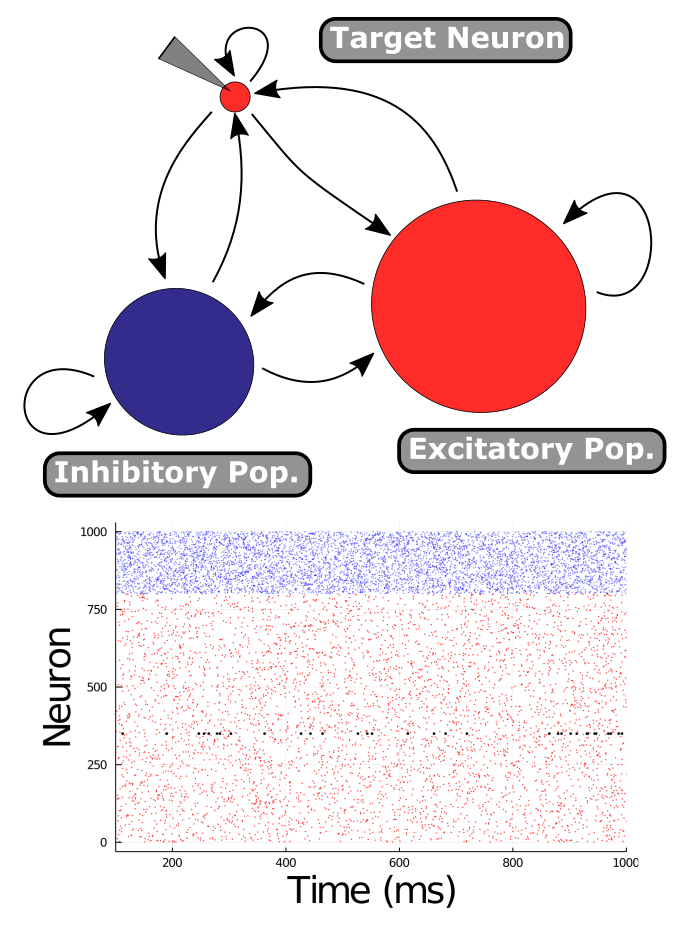}
\caption{\label{fig:Motivation} \textbf{Network model} (A) A graphical representation of the network architecture being studied. (B) An example raster plot generated from an extended network of spiking neurons modeled by Eqn.\ref{eqn:HawkesProc}}
\end{figure}

s
The Gaussian process approximation of the membrane dynamics in Eqn.~\ref{subeqn:HawkesMembr} is obtained by marginalizing out the spiking dynamics from the action in Eqn.~\ref{eqn:full_action} and taking a saddle point approximation of the action around the mean-field solution in Eqn.~\ref{eqn:HP_MembrMF} (see Appendices \ref{sec:GPA_PopAvg} \& \ref{sec:PopAvg_GPA} for details). The resulting action corresponds to the Gaussian stochastic process given by \cite{OckerEtAl2017}
\begin{subequations}
\label{eqn:NetwGPA}
\begin{equation}
    d\mathbf{V} =  \mathbf{A}\left(\mathbf{V}^{mf}-\mathbf{V}\right)dt+\mathbf{\Sigma} d\mathbf{W}_t\label{subeqn:GPA_SDE}
\end{equation}
\begin{align}
       A_{ij} = \delta_{ij}\left(\tm^{-1}  + \ts^{-1}J_{\rm self}\phi'(V_j^{\rm mf})\right) -\ts^{-1}w_{ij}\phi'(V_j^{\rm mf})
\end{align}

\begin{align}
    \left(\mathbf{\Sigma}\mathbf{\Sigma}^T\right)_{ij} = \ts^{-2}\sum_k & \Bigg[ \left(-\delta_{ik}J_{\rm self} + w_{ik} \right)\nonumber\\
    &~~~~\times\left(-\delta_{jk}J_{\rm self} + w_{jk} \right)\phi(V_k^{\rm mf})\Bigg]
\end{align}
\end{subequations}
\normalsize
where $d\mathbf{W}_t$ is a standard Wiener process and we use the It\^{o} convention. We note that Eqn.~\ref{subeqn:GPA_SDE} is an Ornstein-Uhlenbeck (OU) process, albeit one in which the drift and diffusion matrices are dependent on the mean-field values of the membrane potential.

In principal, the network modeled in Eqn.~\ref{eqn:HawkesProc} and approximated in Eqn.~\ref{eqn:NetwGPA} could be of arbitrary size. To make our information geometric analysis tractable, however, we will reduce the model to a three-population model, comprising excitatory and inhibitory populations, and a single neuron targeted with an injected current; this is depicted diagrammatically in Fig.~\ref{fig:Motivation}. We start by considering the connectivity matrix to be random with each entry being a Bernoulli variable with probability $p$ being scaled by a connection type-dependent value $w_{IJ}$. To produce the more tractable reduced model, we take a population-averaging approach to the approximated process in Eqn.~\ref{eqn:NetwGPA}. We now use an uppercase subscript to denote a population averaged variable. For example
\begin{equation*}
     V_I \equiv \frac{1}{N_I}\sum_{i\in I}V_i(t),
\end{equation*}
where we will use uppercase indices $I,J,K \in \{0, 1, 2\}$ to denote the different populations, with $I = 0$ the single test neuron, $I = 1$ the excitatory population, and $I = 2$ the inhibitory population. The dynamics of the population-averaged membrane potentials under the Gaussian approximation now follow a lower-dimensional version of Eqn.~\ref{subeqn:GPA_SDE} with drift and diffusion matrices given by
\begin{subequations}
\label{eqn:PopGPA}
\begin{align}
       A_{IJ} =& \delta_{IJ}(\tau_m^{-1} + \tau_s^{-1} J_{\rm self} \phi'(V_I^{\rm mf}))\nonumber\\
       &~~~~~~~~~~ - \tau_s^{-1} p w_{IJ} N_J \phi'(V^{\rm mf}_J)
\end{align}
\begin{align}
    \left(\mathbf{\Sigma}\mathbf{\Sigma}^T\right)_{IJ} = &\tau_s^{-2}\sum_{K=0,1,2}\Bigg[\left(-\delta_{IK}\frac{J_{\rm self}}{N_K}  +  pw_{IK} \right)\nonumber\\
    &\times\left(-\delta_{JK}\frac{J_{\rm self}}{N_K}  + p w_{JK} \right) N_K \phi(V^{\rm mf}_K)\Bigg]\\
    &\approx \tau_s^{-2} \sum_{K=1,2} p^2 w_{IK} w_{JK} N_K \phi(V^{\rm mf}_K)\nonumber
\end{align}
The approximation in the last line above comes from the fact that $N_1,~N_2\gg 0$. The population-averaged mean-field equations are now 
\begin{align}
    \footnotesize V_I^{\rm mf}=&\varepsilon_I+\tm I_{I}+\frac{\tm}{\ts}\mu_{\rm ext}\nonumber\\
    &+\frac{\tm}{\ts}\left( -J_{\rm self}\phi(V_I^{\rm mf})+\sum_{J=0,1,2}pw_{IJ}N_J\phi(V_J^{\rm mf})\right)\normalsize\label{eqn:Pop_MeanField}
\end{align}
\end{subequations}
We formally derive the Gaussian-process approximation of the full-network (Eqn.~\ref{eqn:NetwGPA}) and the population-averaged approximation (Eqn.~\ref{eqn:PopGPA}) in Appendix~\ref{sec:GPA_PopAvg}. We also note that the statistics of the model in Eqn.~\ref{eqn:PopGPA} are equivalent to those derived by first taking a population average of the membrane potential dynamics and then applying the Gaussian approximation framework. This second derivation is provided in Appendix~\ref{sec:PopAvg_GPA}.

The spiking model we study centers around a balanced network, specifically a network that is not finely tuned. This notion of fine-tuning arises from a standard derivation of balance equations for the model system (see Appendix \ref{sec:BalEqns}). In short, we can look at the average external input $\kappa_I$ into population $I$. For our model, we can approximate $\kappa_I$ to leading order as:
\footnotesize
\begin{align*}
    \tau_s^{-1}\kappa_I \approx \sqrt{N}&\left( \frac{1}{\sqrt{N}}\left(I_{I} +\ts^{-1}\mu_{\rm ext} \right)\right.\\
    &\left.+ \tau_s^{-1} \left\{ pw_{I1} \frac{N_1}{\sqrt{N}} \phi(V_1) + pw_{I2} \frac{N_2}{\sqrt{N}} \phi(V_2) \right\} \right)
\end{align*}
\normalsize
where $V_I$ is the population-averaged membrane potential for population $I$. For the model to be in a balanced state, the variance of the synaptic input should be $\BigO(N^0)$ which in turn implies the synaptic weights should scale as $w_{IJ}\sim 1/\sqrt{N}$. Additionally, we assume that $N_I\propto N$ and $\langle I_{I}\rangle,~\mu_{\rm ext}\propto\sqrt{N}$. The balanced state of the model also requires that all $\kappa_I$ be $\BigO(1)$. For this to be true as $N\rightarrow\infty$, the terms in the parentheses must vanish. This provides gives us a linear system that uniquely defines $\left(\phi(V_1),\,\phi(V_2)\right)$:
\small
\begin{equation}
    -\begin{bmatrix} I_{1} +\ts^{-1}\mu_{\rm ext} \\ I_{2} +\ts^{-1}\mu_{\rm ext}  \end{bmatrix} = \frac{1}{\ts}\begin{bmatrix} pw_{11} N_1  & pw_{12} N_2\\pw_{21} N_1  & pw_{22} N_2  \end{bmatrix}\begin{bmatrix} \phi(V_1)\\ \phi(V_2)  \end{bmatrix}\label{eqn:MeanBalInputSystem}
\end{equation}
\normalsize
From this set of equations, we derive two cases. 
First, if the matrix on the right-hand side of Eqn.~\ref{eqn:MeanBalInputSystem} is singular and neither of the columns of the matrix are trivially the zero-vector, the columns must be scalar multiples of each other. We refer to this as a ``fine-tuned'' spiking model. If the left-hand side of Eqn.~\ref{eqn:MeanBalInputSystem} is also a multiple of the columns, the system admits an infinite set of solutions $\left(\phi(V_1),\,\phi(V_2)\right)$. Otherwise, it admits no solution. Such a network is thus finely-tuned to specific inputs. In contrast to this, we have ``un-tuned'' spiking models. In this case, the matrix on the right-hand side of Eqn.~\ref{eqn:MeanBalInputSystem} is invertible and the system admits a unique solution $\left(\phi(V_1),\,\phi(V_2)\right)$. This in effect applies constraints on the values of $\{w_{IJ}\}$, which we refer to as the balance equations Eqns.~\ref{eqn:balance_conditions1} \& \ref{eqn:balance_conditions2} (see Appendix \ref{sec:BalEqns} for a derivation). Moving forward, we consider only spiking models derived from a balanced, un-tuned network. We also introduce a linear non-spiking model that will serve as a baseline comparison.

\subsection{Linear non-spiking model}

Although the Gaussian process approximation of the spiking network will have a Gaussian steady-state distribution of the membrane potentials, the parameters of this distribution vary nonlinearly with the self-consistent mean-field solutions. To demonstrate that the behaviors we observe are consequences of the mean-field treatment of the spiking network, and not just the behavior of Gaussian processes, we also construct a simpler model of networked, linear non-spiking (or ``graded potential") neurons. We assume the neurons are injected with large numbers of synaptic input that sum together to be approximately Gaussian, with non-zero mean $\mu_{\rm ext}$, creating a stochastic system with dynamics described by:
\begin{align}
    \frac{dV_i}{dt} =& -\tm^{-1}(V_i-\varepsilon_I)+I_i + \ts^{-1}\mu_{\rm ext}-\ts^{-1}J_{\rm self}\phi(V_i)\nonumber\\
    &~~~~~~~~~+\ts^{-1}\sum_j w_{ij}\phi(V_j) + \xi_i(t)
\end{align}

Here, the transfer function $\phi(\cdot)$ is simply the identity function (i.e.~$\phi(x)=x$). The processes $\xi_i(t)$ are zero-mean Gaussian noise synaptic input from neurons external to the network being examined, and thus they scale with $\ts^{-1}$. We define the covariance of the noise processes $\{\xi_i(t)\}$ as follows.
\begin{equation*}
    \langle \xi_i(t)\xi_j(t') \rangle = \ts^{-2}\delta_{ij}\mu_{\rm ext}\delta(t-t')
\end{equation*}

After population-averaging, the non-spiking model becomes another OU process:
\begin{align}
    d\mathbf{V} =&\mathbf{A}\left(\mathbf{A}^{-1}\left(\ts^{-1}\mathbf{\mu}_{\rm ext}+\tm^{-1}\mathbf{\varepsilon}_I+\mathbf{I}\right) - \mathbf{V}\right)dt+\mathbf{\Sigma}d\mathbf{W}_t\nonumber\\
    =&\mathbf{A}\left(\mathbf{\mu} - \mathbf{V}\right)dt+\mathbf{\Sigma}d\mathbf{W}_t.\label{eqn:PopLFRN}
\end{align}
The drift and diffusion matrices are defined as follows
\begin{align*}
    A_{IJ} =&  \delta_{IJ}\tm^{-1}+\ts^{-1}w_{IJ}^\ast\\
    w_{IJ}^\ast &= -\delta_{IJ}J_{\rm self}+ pw_{IJ}^{\rm mod}N_J\\
    \left(\Sigma\Sigma^T\right)_{IJ} &= \ts^{-2}\delta_{IJ}\frac{\mu_{\rm ext}}{N_I}
\end{align*}
Here and in the following sections, $\mathbf{w}^\ast$ denotes the effective connectivity matrix for the linear non-spiking models. The values of $\mathbf{w}^{\rm mod}$ are modulated depending on the desired excitation-inhibition conditions, which will be discussed in Sec.~\ref{sec:Archs}.

The linear form of the population-averaged non-spiking model permits more analytic study than the corresponding spiking models. Ornstein-Uhlenbeck processes like those in Eqns.~\ref{eqn:PopGPA} and \ref{eqn:PopLFRN} admit a Gaussian steady-state distribution if all eigenvalues of the drift matrix are positive \cite{VatiwutipongPhewchean2019}. From the form of the drift matrix for the linear model (Eqn.~\ref{eqn:PopLFRN}) there is a correspondence between eigenvalues of the drift matrix $\mathbf{A}$ and the connectivity matrix $\mathbf{w}^\ast$.
\begin{equation*}
    \lambda_{i,\mathbf{A}}=\tm^{-1}-\ts^{-1}\lambda_{i,\mathbf{w}^\ast}.
\end{equation*}
From the stationarity condition on the eigenvalues of $\mathbf{A}$ and this correspondence between eigenvalues of $\mathbf{A}$ and $\mathbf{w}^\ast$, we can derive a stability boundary for the $(\tm^{-1},\ts^{-1})$ inverse timescale-space
\begin{equation}
    \tm^{-1}>\ts^{-1}\lambda_{\mathbf{w}^\ast},\,\,\,\forall\,\,\lambda_{\mathbf{w}^\ast}.\label{eqn:StabBound}
\end{equation}

The loss of stability observed in OU processes often corresponds to a non-stationary regime in which the random variables may grow without bound. As the firing rate nonlinearity $\phi(x)$ used in the spiking model is quasi-linear in the $x>0$ regime, we expect the stability of the spiking models to be similar to the non-spiking models when they have the same E/I-dependent connectivity given by $w_{IJ}$.

\begin{table*}[t]
\centering
\begin{tabular}{|p{1.47cm}||p{11.5cm}|p{3.5cm}||} 
 \hline
 Parameter & Description & Value\\ [0.5ex] 
 \hline\hline
  $N$ & Total number of neurons & 1000 \\ \hline
 $N_e$ & Number of excitatory neurons & $0.8N$ \\ \hline
 $N_i$ & Number of inhibitory neurons & $0.2N$ \\\hline
 $p$ & Probability of a directional synaptic connection $w_{ij}$ between any two neurons & 0.1 \\ \hline
 $-J_{{\rm self}}$ & The self connection for a neuron of type $E/I$ designed to capture post-spike refractory dynamics & $-5$ \\\hline
 $\varepsilon_I$ & Leak reversal potential for neuron $i$ & 0 \\ \hline
 $I_{I}$ & Injected current impinging on neuron population $I$ & \footnotesize$\begin{cases}
     0.02\,\,{\rm if \, target}\\0\,\,{\rm otherwise} 
 \end{cases}$\normalsize \\ \hline
 $\mu_{\rm ext}$ & Mean input from network-external neurons & $0.1$ \\ \hline
 $w_{ee,{\rm base}}$ & The total expected synaptic input weight from exc. neurons to exc. neurons & $285$ \\ \hline
 $w_{ie,{\rm base}}$ & The total expected synaptic input weight from exc. neurons onto inh. neurons & $300$ \\ \hline
 $w_{ei,{\rm base}}$ & The total expected synaptic input weight from inh. neurons to exc. neurons & $-902.5$ \\ \hline
 $w_{ii,{\rm base}}$ & The total expected synaptic input weight from inh. neurons to inh. neurons & $-950$ \\ \hline
 $\phi(x)$ & Firing rate transfer function & \footnotesize$\begin{cases}
     x\,\,{\rm if \, Non-spiking}\\ \Bigg.\frac{x+\sqrt{x^2+\frac{1}{2}}}{2}\Bigg.\,\,{\rm if \, Spiking} 
 \end{cases}$\normalsize\\ \hline
 $\tm$ & Membrane timescale & variable \\\hline
 $\ts$ & Synaptic timescale & variable \\ [1ex] \hline
\end{tabular}
\caption{\textbf{Model parameters} Descriptions and numerical values for the parameters for the non-spiking and spiking model types.}
\label{tbl:ModelParams}
\end{table*}

\subsection{Stationary distributions\label{sec:StatDist}}
As mentioned above, the stationary distributions admitted by Ornstein-Uhlenbeck processes are Gaussian when they exist \cite{VatiwutipongPhewchean2019}. Consider a general $N$-dimensional OU process:
$$d\mathbf{X} = \mathbf{A}\left(\mathbf{\mu}-\mathbf{X}\right)dt +\mathbf{\Sigma}d\mathbf{W}_t$$
The stationary distribution, when it exists, is described by the multivariate normal probability density \cite{VatiwutipongPhewchean2019}
$$p\left(\mathbf{X}\right) =\frac{1}{(2\pi)^{N/2} \sqrt{{\rm det}(\mathbf{C})}}e^{- \frac{1}{2} (\mathbf{X}-\mathbf{\mu})^T \mathbf{C}^{-1} (\mathbf{X}-\mathbf{\mu})}, $$
where the stationary covariance $\mathbf{C}$ is given by the solution to the matrix equation \cite{VatiwutipongPhewchean2019}
$$\mathbf{\Sigma}\mathbf{\Sigma}^{T} = \mathbf{A}\mathbf{C}+\mathbf{C}\mathbf{A}^T.$$
In practice, the stationary covariance matrix can by found by linearizing the matrix equation and solving the resulting linear system numerically.
\subsection{Network architectures}\label{sec:Archs}

Now, we turn back to our network models. We consider a population of excitatory and inhibitory neurons in which a single excitatory target neuron is injected with an external driving current.  The full network contains $N=1000$ sparsely connected neurons. We condensed the full network model into representative 3-neuron network by population-averaging, as depicted in Fig.\ref{fig:Motivation}A and described in Eqn.~\ref{eqn:PopGPA}, representing the excitatory target neuron, the excitatory population, and the inhibitory population. Table \ref{tbl:ModelParams} contains descriptions and numerical values for the parameters used in the present study.

In addition, we would like to adjust the relative recurrent excitation and inhibition in the networks. To accomplish this, the base connection weights given in Table \ref{tbl:ModelParams} are scaled by a ratio $r>0$ depending on the desired activity regime: 
\begin{subequations}
    \begin{equation}
        \mathbf{w}_{Xe}^{\rm mod}=r_e(r)w_{Xe,{\rm base}}=\begin{cases}
            rw_{Xe,{\rm base}}~~~{\rm if}~r\geq 1\\
            w_{Xe,{\rm base}}~~~{\rm otherwise}
        \end{cases}
    \end{equation}
    \begin{equation}
        \mathbf{w}_{Xi}^{\rm mod}=r_i(r)w_{Xi,{\rm base}}=\begin{cases}
            w_{Xi,{\rm base}}~~~{\rm if}~r\geq 1\\
            \frac{1}{r}w_{Xi,{\rm base}}~~~{\rm otherwise}
        \end{cases}.\end{equation}\label{eqn:WBaseScaling}\\
\end{subequations}
The ratio serves to boost the recurrent excitatory weights in the excitatory regime ($r>1$) and the recurrent inhibitory weights in the inhibitory regime ($r<1$), through the functions $r_e(r)$ and $r_i(r)$, respectively. 

The connection matrices $\mathbf{w}$ and $\mathbf{w}^\ast$ of the population-averaged spiking and non-spiking models, respectively, are now constructed from the full-network parameters and scaling of excitation and inhibition. All matrices $\mathbf{w}$ and $\mathbf{w}^\ast$ use the same indexing with $I=0$ denoting the target neuron ``population,'' $I=1$ denoting the remaining excitatory neurons, and $I=2$ denoting all inhibitory neurons. The connection strengths used in the linear non-spiking model in Eqn.~\ref{eqn:PopLFRN} is then given by

\begin{widetext}
     \begin{align}
        \mathbf{w}^\ast =& -\begin{bmatrix} J_{{\rm self}} &0 &0\\0 &J_{{\rm self}} & 0\\ 0& 0& J_{{\rm self}} \end{bmatrix} + \frac{1}{\sqrt{pN}}\begin{bmatrix} pw_{ee} & p\left(N_e-1\right)w_{ee} & pN_iw_{ei}  \\ pw_{ee} & p\left(N_e-1\right) w_{ee} & pN_iw_{ei}\\pw_{ie} & p\left(N_e-1\right)w_{ie} &  pN_iw_{ii}        \end{bmatrix}\nonumber\\
        &= -\begin{bmatrix} J_{{\rm self}} &0 &0\\0 &J_{{\rm self}} & 0\\ 0& 0& J_{{\rm self}} \end{bmatrix} + \frac{1}{\sqrt{pN}}\begin{bmatrix} pr_e(r)w_{ee,{\rm base}} & pr_e(r)\left(N_e-1\right)w_{ee,{\rm base}} & pr_i(r)N_iw_{ei,{\rm base}}  \\ pr_e(r)w_{ee,{\rm base}} & pr_e(r)\left(N_e-1\right) w_{ee,{\rm base}} & pr_i(r)N_iw_{ei,{\rm base}}\\pr_e(r)w_{ie,{\rm base}} & pr_e(r)\left(N_e-1\right)w_{ie,{\rm base}} &  pr_i(r)N_iw_{ii,{\rm base}}        \end{bmatrix}. \label{eqn:W_NSpk}
    \end{align}
\end{widetext}
The $1/\sqrt{pN}$ scaling of the connection weights arises from the balance conditions mentioned at the end of Sec.~\ref{sec:NLHP} and derived in Appendix \ref{sec:BalEqns}. The connection matrices used by the spiking models described generally in Eqn.~\ref{eqn:PopGPA} is given by
\begin{align}
    \mathbf{w} &= \frac{1}{\sqrt{pN}} \mathbf{w}^{\rm mod}\nonumber\\
    &=\frac{1}{\sqrt{pN}}\begin{bmatrix} r_e(r)w_{ee,{\rm base}} & r_e(r)w_{ee,{\rm base}} & r_i(r)w_{ei,{\rm base}}  \\ r_e(r)w_{ee,{\rm base}} & r_e(r)w_{ee,{\rm base}} & r_i(r)w_{ei,{\rm base}}\\r_e(r)w_{ie,{\rm base}} & r_e(r)w_{ie,{\rm base}} &  r_i(r)w_{ii,{\rm base}}        \end{bmatrix}.\label{eqn:W_Spk}
\end{align}

Finally, we would like a measure of the balance of excitation and inhibition (``E/I") within a class of models. As each model type corresponds to many particular models with different values of the inverse timescales $(\tm^{-1},\ts^{-1})$, we require a proxy measure for the E/I ratio to describe the whole class. In line with the method for adjusting the relative strength of recurrent excitation and inhibition introduced above, we assign a ratio of connection weights into the bulk excitatory population for a given model and a given modulation $r$. For the non-spiking models, we give the log-ratio $R$ of these weights
\begin{equation*}
    R = \log_{10}
    \left|\frac{\mathbf{w}^\ast_{ 2,1} +\mathbf{w}^\ast_{2,2}}{\mathbf{w}^\ast_{2,3}}\right|
\end{equation*}
To make an accurate comparison to the non-spiking models, the E/I values for the spiking models are reported using this same measure (for a given value of modulation parameter $r$). 

We note here that the same balanced-network calculations (see Appendix \ref{sec:BalEqns}) that gave rise to the definitions of ``fine-tuned'' and ``un-tuned'' formally define a notion of balance. A balanced spiking network based on the model architecture used here must satisfy constraints on the weights of $\mathbf{w}$ (Eqn.~\ref{eqn:W_Spk}), either Eqns.~\ref{eqn:balance_conditions1} or Eqns.~\ref{eqn:balance_conditions2}. The base connection weights for the unadjusted network---i.e.~$r=1$ in Eqns.~\ref{eqn:WBaseScaling}---were chosen to meet these balance criteria, and the functions $r_e(r)$ and $r_i(r)$ serve to tilt the excitation-inhibition balance with respect to this measure.

\subsection{Timescale sampling}\label{sec:TimescaleSampling}

\begin{figure}[t]
\includegraphics[width=0.8\linewidth]{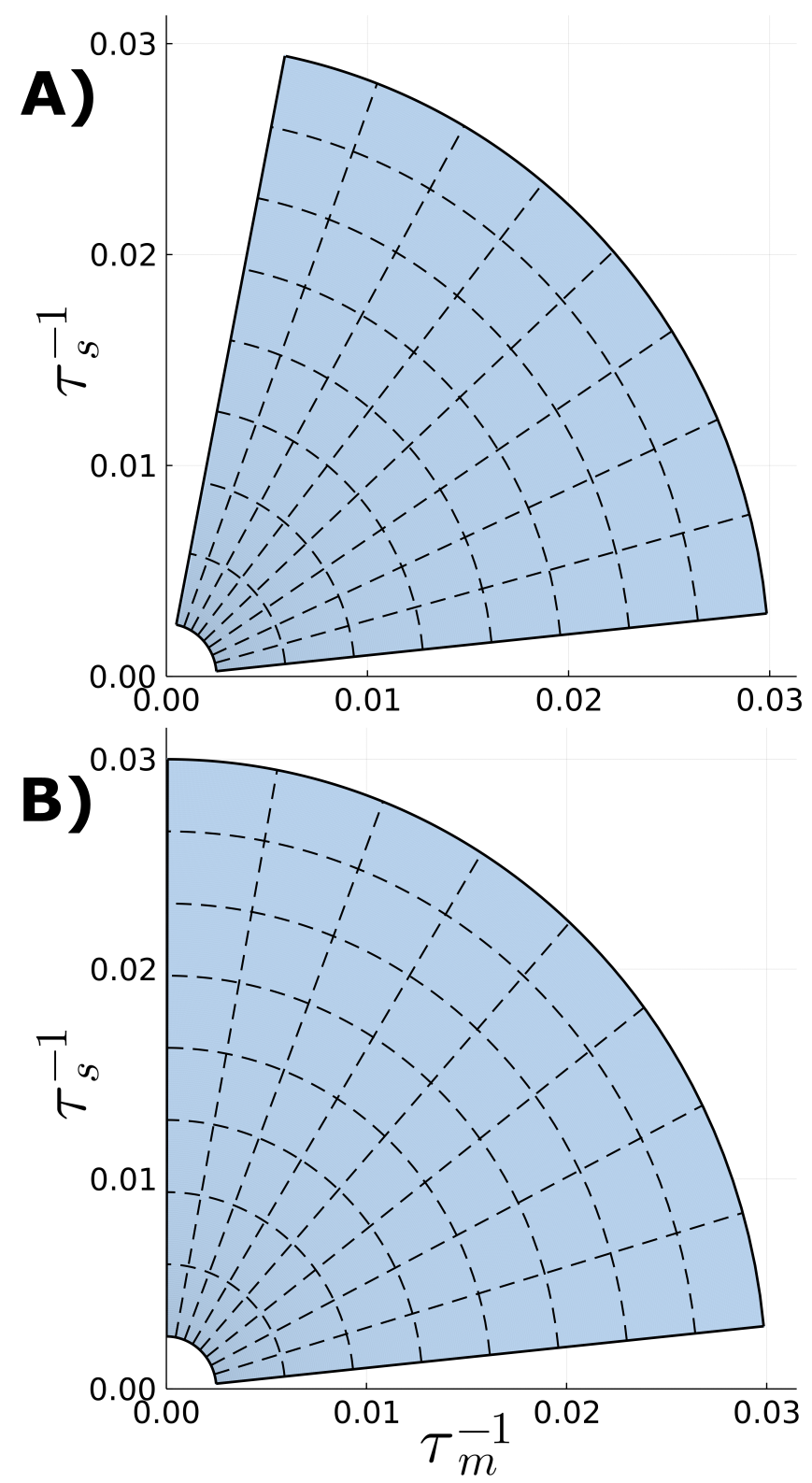}
\caption{\label{fig:Sampling} \textbf{Maximal sampling of the inverse timescale plane:} We sample pairs of inverse-timescale values from the depicted region of the $\tau^{-1}_m$-$\tau^{-1}_s$ plane. Sampled points are distributed evenly on a curvilinear grid between a predefined lower bound and the stability boundary for the specific connection matrix being used (A). If sampling to the stability boundary would produce samples with negative timescales, the space is instead sampled up to a predefined value in the angular direction (B).}
\end{figure}

To embed and visualize the model manifolds of interest, we must sample points on the manifold characterized by different values of the two modulated parameters. We do this by sampling a portion of the inverse-timescale parameter space that satisfies the stability condition given by Eq.~(\ref{eqn:StabBound}) and where both inverse-timescales are positive. We apply a curvilinear grid to this region, uniformly sampling the radial and angular components.
The radial distance components $d$ of the grid are taken over a fixed range:
\begin{equation*}
    d\in [0.0025,\,0.03] \,\,{\rm ms}^{-1}
\end{equation*}
To apply both the stability boundary and positivity constraints, the lower bound of the angular component $\alpha$ of the sample grid is set to a fixed value while the upper bound is set either by the stability boundary described by Eqn.~\ref{eqn:StabBound} or to a fixed value, whichever is more stringent:
\begin{equation*}
    \tan(\alpha)\in \left[0.1,\,\min\left(\frac{1}{\max\{\lambda_{\mathbf{w}^\ast}\}},500\right)\right]
\end{equation*}
The conditions for this maximal sampling are summarized in Table~\ref{tbl:Sampling}. It is important to note that the stability boundary is determined by the eigenvalues of the connectivity matrix $\mathbf{w}^\ast$, and thus the stability boundary and the sampling region are affected by the induced E/I balance is adjusted through its affect on $\mathbf{w}^\ast$. The spiking models use the connection matrix from the equivalent non-spiking model to set the sampling range. The maximal sampling scheme is depicted diagrammatically in Fig.~\ref{fig:Sampling}.

After the maximal sampling of parameter space for each model type for each E/I condition, sample points from the inverse-timescale space are subject to further exclusionary criteria. 
For both the spiking- and non-spiking-type models, sample points are excluded if they cause either the drift matrix $\mathbf{A}$ or the covariance matrix $\mathbf{C}$ to become singular. The singularities in these matrices have been observed to occur numerically near the theoretical stability boundary (Eqn.~\ref{eqn:StabBound}). 
In addition, sample points for the spiking models are excluded if the Euler integration used to find the mean-field solutions to Eqn.~\ref{eqn:Pop_MeanField} does not converge. The integration is determined to be numerically non-convergent if the rate of change of the system either exceeds a predetermined value during integration or does not hit a convergence threshold before reaching the maximum number of steps. 

All model manifolds studied here were generated from between 210,000 and 211,000 sampled parameter pairs.

\begin{table}[t]
\centering
\begin{tabular}{|m{1.7cm}||m{2.5cm}|m{2.9cm}||} 
 \hline
  & Radial distance $d$ & Angle $\alpha$ \\ [0.5ex] 
 \hline\hline
 Minimum Value & $0.0025 \,\,{\rm ms}^{-1}$ & $0.1$ \\ \hline
 Maximum Value & $0.03 \,\,{\rm ms}^{-1}$ & $\min\left(\frac{1}{\max\{\lambda_{\mathbf{w}^\ast}\}},500\right)$ \\ \hline
 Number of Samples & $301$ & $701$  
  \\[1ex] 
 \hline
\end{tabular}
\caption{\textbf{Maximal Sampling Parameters} Descriptions and numerical values for the parameters that are constant across the non-spiking, spiking with fine-tuning, and spiking without fine-tuning model types.}
\label{tbl:Sampling}
\end{table}

\section{isKL Embedding}\label{sec:InfoGeo}
In this section, we recapitulate the methods developed by Teoh and colleagues \cite{TeohEtAl2020}. This framework revolves around using the symmetric Kullback-Liebler divergence $D_{sKL}$ as a measure of separation for different probabilistic models of the same form but with different parameters: 
\begin{align*}
    D_{sKL}(\theta,\theta')=& D_{KL}(\theta:\theta') + D_{KL}(\theta':\theta)\\ &=\mathbb{E}_{\theta}\left[\ln{\frac{p(x|\theta)}{p(x|\theta')}}\right]-\mathbb{E}_{\theta'}\left[\ln{\frac{p(x|\theta)}{p(x|\theta')}}\right]. \label{eqn:DsKLDef}
\end{align*}
Teoh \emph{et al.} apply this measure to exponential family models, which have the general form 
\begin{equation*}
p(x|\theta) = \exp\left[\sum_{i=1}^nt_i(x)\eta_i\left(\theta\right)+ k(x)- A\left(\eta\left(\theta\right)\right)\right], \label{eqn:ExpFamDef}
\end{equation*}
where $\{\eta_i\left(\theta\right)\}$ are the $n$ natural parameters of the model and $\{t_i(x)\}$ are the corresponding sufficient statistics. The $D_{sKL}$ for exponential family models can be analytically decomposed into a \emph{finite} number of component functions
\small
\begin{equation*}
    D_{sKL}[\theta,\theta'] = \sum_{i=1}^n\left\{ \left[\mathcal{T}_i^+(\theta)-\mathcal{T}_i^+(\theta')\right]^2-\left[\mathcal{T}_i^-(\theta)-\mathcal{T}_i^-(\theta')\right]^2\right\}.
\end{equation*}
\normalsize
These component functions form a set of $n$ space-like ($\mathcal{T}_i^+$) and $n$ time-like ($\mathcal{T}_i^-$) coordinates by which the model manifold may be embedded in a Minkowski-like behavioral space \cite{TeohEtAl2020}. These coordinate functions are given in terms of the natural parameters and sufficient statistics \cite{TeohEtAl2020} by
\begin{equation*}
    \mathcal{T}_i^{\pm} = \frac{1}{2}\left[\eta_i\left(\theta\right)\pm\langle t_i(x)\rangle_{\theta}\right].
\end{equation*}
Alternatively, we may use an isometric embedding given by shifting and rotating the manifold \cite{TeohEtAl2020}:
\begin{equation}
    T_i^{\pm}(\theta) = \frac{1}{2}\left\{\lambda_i\left[ \eta_i(\theta)-\overline{\eta_i}\right]\pm\frac{1}{\lambda_i}\left[\langle t_i\rangle_\theta-\overline{\langle t_i\rangle}\right] \right\}.
\end{equation}
We use $T^\pm$ to distinguish the isometric embedding coordinates from the unscaled coordinates $\mathcal T^\pm$. Here, an over-bar denotes a mean over sampled parameters and $\lambda_i = \left[\var\left(\langle t_i\rangle\right)/\var\left(\eta_i\right)\right]^{1/4}$. These coordinates can be understood as an alternative definition of the exponential family. We can straightforwardly express the log-likelihood function for an exponential family in terms of the isKL coordinates:
\begin{align*}
    \ln p(x|\theta) =& \ln k(\mathbf{x})+\sum_i\mathcal{T}_i^+(\theta)t_i(\mathbf{x})\\
    &\,\,\,\,\,\,\,\,\,\,+\sum_i\mathcal{T}_i^-(\theta)t_i(\mathbf{x})-A(\theta)\\
    &= g(\mathbf{x})+\sum_iT_i^+(\theta)t_i(\mathbf{x})\\
    &\,\,\,\,\,\,\,\,\,\,+\sum_iT_i^-(\theta)t_i(\mathbf{x})-A(\theta),
\end{align*}
where $g(\mathbf{x})=\ln k(\mathbf{x})+\sum_i\overline{\eta}_it_i(\mathbf{x})$. The authors \cite{TeohEtAl2020} show that the coordinates $\{T_i\}$ can also be understood in relation to the data visualization procedure  multidimensional scaling (MDS). In standard MDS the data points are recorded data, whereas here each ``data point'' corresponds to the full distribution of an exponential family evaluated at a specific set of parameters. The double mean centered matrix of MDS can be constructed in this context from the pairwise separation matrix measured by the symmetric KL-divergence $\mathbf{D}_c = -\mathbf{P}\mathbf{D}_{sKL}\mathbf{P}$ with $\mathbf{P}_{ij}=1/n - \delta_{ij}$. In the continuous sampling limit, the eigenvalue problem for MDS is formulated as an integral equation:
\begin{equation}
    \int D_c(\Tilde{\theta},\theta)v(\theta)d\rho(\theta) = \Lambda v(\Tilde{\theta}),
    \label{eqn:Dskleigen}
\end{equation}
where $d\rho(\theta) = \rho(\theta)d\theta$ is the measure of the distribution of parameters $\theta$. Teoh and colleagues show \cite{TeohEtAl2020} that the coordinates $T_i^{\pm}$ are solutions of this eigenvalue problem with corresponding eigenvalues 
\begin{equation}
    \Lambda_i^{\pm} =\frac{1}{2}\left[{\rm Cov}(\eta_i,\langle t_i\rangle)\pm\sqrt{\var(\eta_i)\var(\langle t_i\rangle)}  \right]\label{eqn:coordEVs}
\end{equation}
This procedure produces an embedding with only a finite and relatively small number of non-zero modes, contrasting sharply with the infinite or data-proportional embedding produced by other methods for continuous or discrete parameter sampling, respectively \cite{TeohEtAl2020}.

We complement this perspective by viewing this embedding procedure as an eigenmode expansion of the conditional probability $p(x|\theta)$ around the marginalized distribution $p(x)$ for a given prior on the parameters $\theta$:
\begin{equation}
    p(x|\theta) = p(x)+\sum_{i}c_i^+(x)T_i^+(\theta) + \sum_{i}c_i^-(x)T_i^-(\theta)
    \label{eqn:pdfexpansion}
\end{equation}
By defining the inner product of functions on $\Theta$ as 
$$\langle f(\theta), g(\theta)\rangle =\int d\rho(\theta) f(\theta)g(\theta),$$
the modes $\sqrt{\mu(\theta)}v(\theta)$ of Eq.~(\ref{eqn:Dskleigen}) can be shown to be orthogonal as long as the corresponding eigenvalues are distinct. Thus, the coordinate functions $T_i^\pm(\theta)$ are orthogonal with respect to the weight $\rho(\theta)$. Taking advantage of this orthogonality of the coordinate functions, it follows
\begin{align*}
    &\int d\rho(\theta)p(x|\theta)T_j^\pm(\theta) \\
    &= p(x)\int d\rho(\theta)T_j^\pm(\theta) +\sum_{i,\pm} \int d\rho(\theta) c_j^\pm(x)T_j^\pm(\theta)T_i^\pm(\theta)\\
    &=c_j^\pm(x)\int d\rho(\theta)\left(T_j^\pm(\theta)\right)^2
\end{align*}
The first term on the right-hand side vanishes because the mean of each coordinate function is zero by construction, while only the $i = j$ term from the sum survives due to the orthogonality. Thus, we may calculate the coefficient functions $c_i^\pm(x)$ as
\begin{equation*}
    c_i^\pm(x) = \langle T_i^\pm , T_i^\pm\rangle^{-1}\int d\rho(\theta)p(x|\theta)T_i^\pm(\theta).
\end{equation*}

\begin{figure*}[t]
\includegraphics[width=\linewidth]{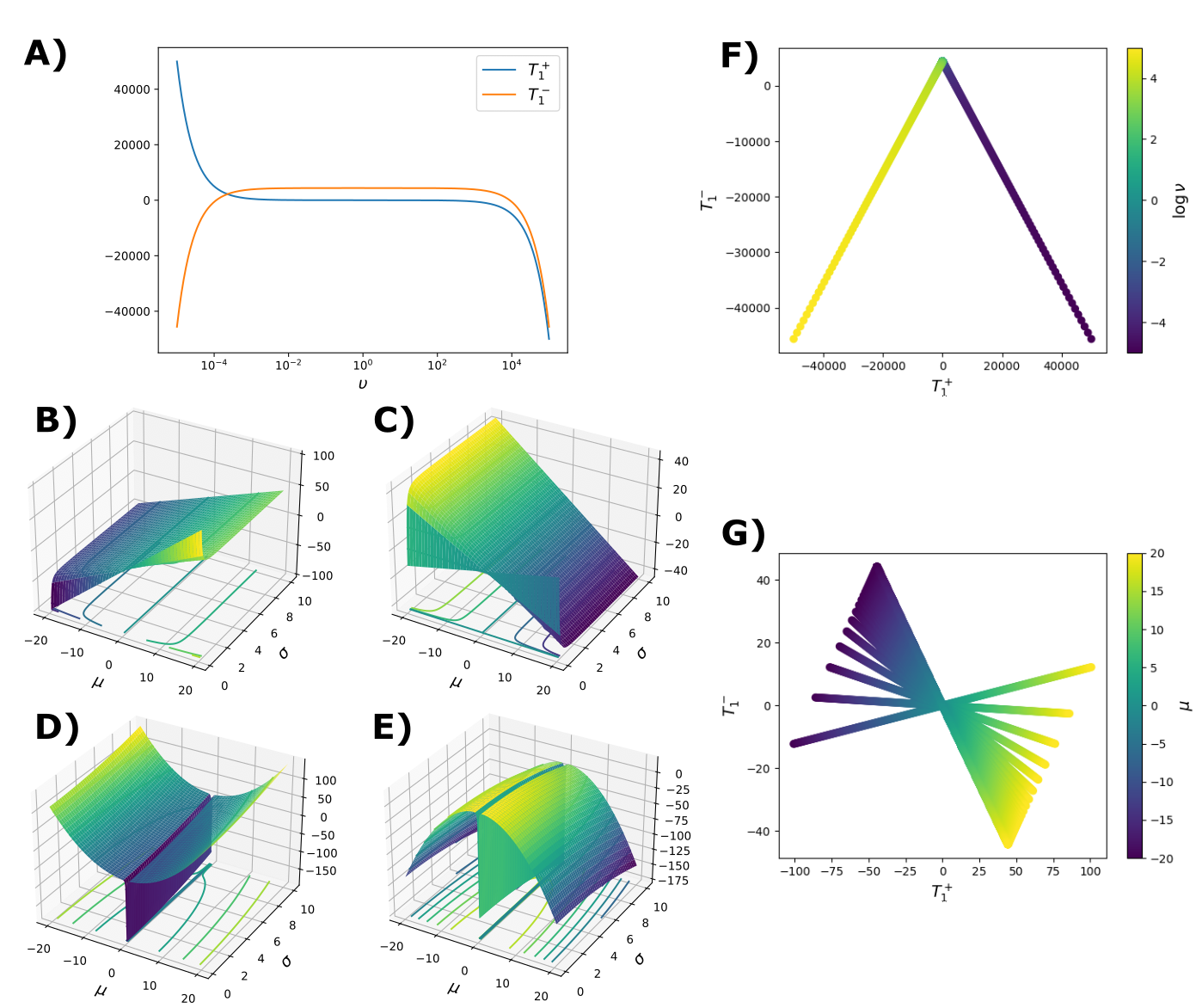}
\caption{\label{fig:CoordVis} \textbf{Visualizations of the isKL embedding coordinate functions:} (A) The two coordinate functions for the 1-dimensional exponential model. (B-E) Coordinate functions for the 1-dimensional Gaussian model. (B) $T_1^+$. (C) $T_1^-$. (D) $T_2^+$. (E) $T_2^-$. (F) The model manifold for the 1-dimensional exponential model colored by $\log{\upsilon}$. (G) The model manifold for the 1-dimensional Gaussian model projected onto just $T_1^\pm$, colored by $\mu$.}
\end{figure*}
In this work we focus on applying these embedding methods to the stationary distributions of the various network models, both multivariate normal within our mean-field approximation. For a $M$-dimensional multivariate normal distribution with a set of means $\{\mu_i\}$ and covariance values $\{C_{ij}\}$, the $M(M+3)/2$ distinct natural parameters and sufficient statistics are given by
\begin{equation}
    \eta = \begin{bmatrix} \sum_i C^{-1}_{1i} \mu_i\\ \vdots \\ \sum_i C^{-1}_{Mi} \mu_i \\ -\frac{1}{2}C^{-1}_{11} \\ \vdots \\ -\frac{1}{2}C^{-1}_{M1}\\ -\frac{1}{2}C^{-1}_{22} \\\vdots \\-\frac{1}{2}C^{-1}_{MM}  \end{bmatrix},\,\,\,\,\,\left\langle t_i \right\rangle_{\theta} =\begin{bmatrix} \left\langle x_1 \right\rangle\\ \vdots \\ \left\langle x_M \right\rangle \\ \left\langle x_1^2 \right\rangle\\ \vdots \\ \left\langle x_Mx_1 \right\rangle\\\left\langle x_2^2 \right\rangle \\ \vdots \\ \left\langle x_M^2 \right\rangle \end{bmatrix}\label{eqn:natparams_suffstats}
\end{equation}
Before we present the embedding and analysis of the models from Section \ref{sec:Models}, we provide two simpler models as illustrative examples that are related to the Poissonian and Gaussian characteristics of our model.

\subsection*{Example: 1-dimensional exponential model}
Let $X$ be exponentially distributed with rate $\upsilon$, i.e. $X\sim{\rm Exp}(\upsilon)$. In the exponential family formalism, we have
$$\eta=-\upsilon,\,\,\,\,\langle t_i\rangle_\theta=\langle x\rangle_\upsilon= \upsilon^{-1},\,\,\,\, k(x)=1,\,\,\,\,A(\upsilon)=-\ln\upsilon.$$
The isKL embedding coordinates for this model are one-dimensional functions given by 
$$T^\pm(\upsilon) = \frac{1}{2}\left\{\lambda\left[ \overline{\upsilon}-\upsilon\right]\pm\frac{1}{\lambda}\left[\upsilon^{-1}-\overline{\upsilon^{-1}}\right] \right\}.$$
These embedding functions are shown in Fig.~\ref{fig:CoordVis}A using a parameter distribution $\rho(\upsilon) = \left(8\upsilon\ln{10}\right)^{-1}$ with support $\upsilon\in [10^{-5},10^5]$ for illustration.

We may also explicitly calculate the coefficients $c^{\pm}(x)$ for this example,
\begin{align*}
    c^{\pm}(x) =& \langle T^\pm , T^\pm \rangle^{-1}\Bigg[\left(\frac{\lambda}{2}\overline{\upsilon} \mp \frac{1}{2\lambda}\overline{\upsilon^{-1}} \right)\left(-\frac{dZ}{dx}\right) \\
    &-\frac{\lambda}{2}\frac{d^2Z}{dx^2}\pm \frac{1}{2\lambda }Z \Bigg],
\end{align*}
where $Z(x) \equiv \int e^{-v x}d\rho(\upsilon)$ is the moment-generating function of the distribution $\rho(\upsilon)$ with source $-x$. The full model manifold is depicted in Fig.~\ref{fig:CoordVis}F, where points are colored by the logarithm of the rate parameter $\upsilon$. Here, we see the manifold is neatly broken into two branches corresponding to a low event-rate ($\log \upsilon<0$) and a high event-rate ($\log \upsilon>0$).

\subsection*{Example: 1-dimensional Gaussian model}
Let $X$ be normally distributed as $X\sim\mathcal{N}(\mu,\sigma)$. In the exponential family formalism, we have
\begin{align*}
     \eta = \begin{bmatrix} \mu/\sigma^2\\ -\sigma^{-2} \end{bmatrix},\,\,\,\,\,\left\langle t_i \right\rangle_{\theta} =\begin{bmatrix} \left\langle x \right\rangle\\  \left\langle x^2 \right\rangle \end{bmatrix}=\begin{bmatrix}\mu \\ \sigma^2+\mu^2 \end{bmatrix}\\
     k(x)=\frac{1}{\sqrt{2\pi}},\,\,\,\,\,A(\mu,\sigma)=\frac{\mu^2}{2\sigma^2}+\ln\sigma.
\end{align*}

The isKL embedding coordinates are then two-dimensional functions
$$T_1^\pm(\mu,\sigma)  = \frac{1}{2}\left\{\lambda_1\left[ \frac{\mu}{\sigma^2}-\overline{\left(\frac{\mu}{\sigma^2}\right)}\right]\pm\frac{1}{\lambda_1}\left[\mu-\overline{\mu}\right] \right\}$$
\begin{align*}
    T_2^\pm(\mu,\sigma)=  \frac{1}{2}&\left\{\lambda_2\left[ -\frac{1}{\sigma^2}+\overline{\left(\frac{1}{\sigma^2}\right)}\right]\right.\\
    &\,\,\,\,\,\pm\frac{1}{\lambda_2}\left[\sigma^2+\mu^2-\overline{\left(\sigma^2+\mu^2\right)}\right] \Bigg\}.
\end{align*}
The Gaussian model coordinate functions are depicted in Fig.~\ref{fig:CoordVis}B-E using a parameter distribution
\begin{equation*}
    \rho(\mu,\sigma) = \begin{cases}
        1/800\,\,{\rm if}\,\,-20\leq\mu\leq20,\,\,0<\sigma\leq20\\
        0\,\,\,\, {\rm otherwise}
    \end{cases}.
\end{equation*}
A projection of the model manifold onto the space-like and time-like coordinates corresponding to first moment of the model is depicted in Fig.~\ref{fig:CoordVis}G. The points on this projection are colored by the mean parameter $\mu$. Here, we see a degree of rotational symmetry in the manifold projection, separated into negative mean values on the left and positive mean values on the right of the $T_1^+$ center-line. Also note that there are apparent breaks in this manifold projection. These breaks do not reflect a true discontinuity in the structure of the manifold, but instead reflect the density with which the $(\mu,\sigma)$-space is sampled. We will see manifold breaks related to the sampling density in our results for the network models.

\section{Results}\label{sec:Results}
Before proceeding with results, it is helpful to briefly summarize the goal of this paper and the workflow constructed in prior sections. We wish to study the population-averaged behavior of stochastic spiking models as we vary synaptic and membrane timescales, repeating this across a range of relative excitation and inhibition. To do this, we approximate the full spiking network dynamics as a population-averaged multivariate Gaussian process (Eqn.~\ref{eqn:PopGPA}). We choose a sample of inverse timescales as discussed in Sec.~\ref{sec:TimescaleSampling}, and in particular constrain the sampling based on the stability condition for the corresponding non-spiking model (Eqn.~\ref{eqn:StabBound}). Within this sampled regime of timescales, the Gaussian process approximations should be mean-reverting and thus reach a stationary Gaussian distribution. We numerically solve for the vector-mean and the covariance matrix of the stationary Gaussian distribution at each sampled timescale-point. Finally, we embed this manifold of stationary Gaussian distributions into a behavioral space using the isKL methods introduced in Sec.~\ref{sec:InfoGeo}. With the workflow summarized, we may proceed.

\subsection{Gaussian process approximations are stable}\label{sec:GPA_DriftEigs}
A key step in the analysis workflow is to find the stationary distribution for the approximated processes at each sampled timescale-point. For the stationary distribution of an Ornstein-Uhlenbeck process to exist, all of the eigenvalues $\lambda_\mathbf{A}$ of the drift matrix $\mathbf{A}$ must have a positive real component. Basing the upper sampling boundary on the theoretical stability boundary of the related linear model, as well as the check for singularities in the drift matrices $\mathbf{A}$ and covariance matrices $\mathbf{C}$, should ensure this requirement is met. We confirm this by explicitly examining the eigenvalues of the sampled models.

\begin{figure*}[t]
\includegraphics[width=0.95\linewidth]{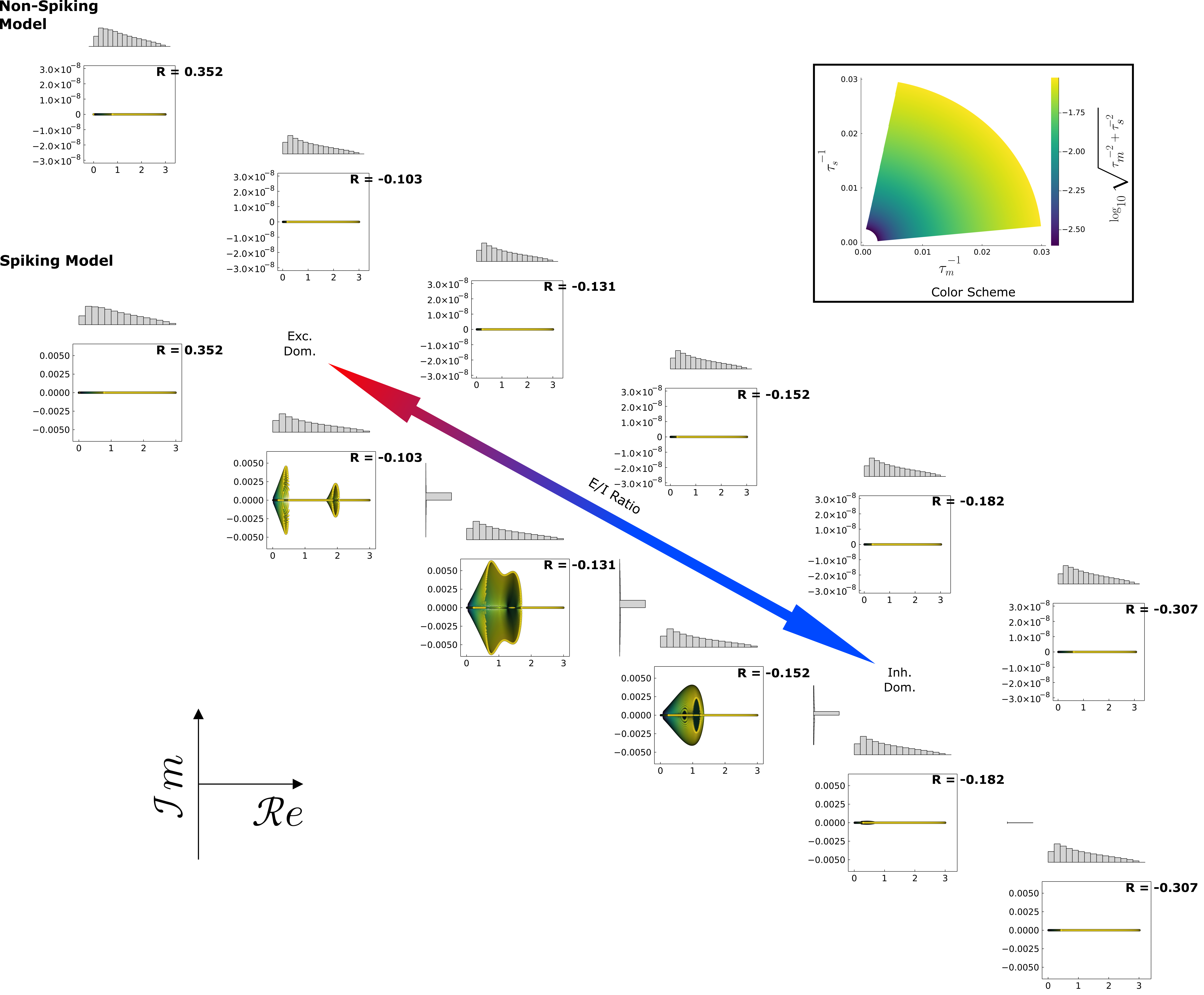}
\caption{\label{fig:MeanRevert_GPA} \textbf{Eigenvalues of drift matrices} Eigenvalue distributions for the drift matrices $\mathbf{A}$ of all sampled network models for the non-spiking model manifolds (top) and the spiking model (bottom) as the excitation-inhibition ratio $R$ is adjusted. Points are colored by the radial distance $d$ of model from the origin in inverse-timescale space, as illustrated by the inset. The marginal histograms for the real ($\mathcal{RE}$, top) and imaginary ($\mathcal{IM}$, right) are given for each distribution. Plots without a histogram in the imaginary dimension indicates a marginal $\delta$-distribution. Emphasis is placed on a portion of the inhibitory regime ($-0.19<R<-0.09$, middle columns) in which some eigenvalue distributions display imaginary components.}
\end{figure*}

Each individual model---as specified by the model type, E/I log-ratio $R$, and a pair of inverse timescales $(\tm^{-1},\ts^{-1})$---has three drift-matrix eigenvalues. We pool together the eigenvalues of all particular models on a given model manifold as specified just by the model type and E/I log-ratio $R$. The resulting eigenvalue distributions for a subset of E/I conditions $R$ are given in Fig.~\ref{fig:MeanRevert_GPA}. Points in the eigenvalue distribution are colored by the log-distance of the particular model (specified by $(\tm^{-1},\ts^{-1})$) from the origin in the inverse-timescale parameter space, shown in the insert. Each individual eigenvalue distribution is accompanied by marginal histograms where appropriate.

The subset of manifolds shown in Fig.~\ref{fig:MeanRevert_GPA} highlights a portion of the inhibition-dominated regime for which the models were observed to have complex eigenvalues. These complex distributions seen in the spiking-type model manifolds have a relatively small range in the imaginary direction and the imaginary components tend to pool near the origin in the along the real-axis. Additionally, most of the density for these complex distributions are along the real-axis itself, indicating that models with complex eigenvalues are relatively rare within their corresponding manifolds. The manifolds for the remaining E/I conditions have eigenvalue distributions qualitatively very similar to those at the extremes: purely real and positive eigenvalues spanning roughly the same range and skewed toward the origin. A key takeaway from Fig.~\ref{fig:MeanRevert_GPA} is that all of the eigenvalues have strictly-positive real components, confirming that the sampled models for each manifold are stable and therefore appropriate for embedding analysis.

\subsection{Behavior of full spiking network models}
As discussed in Sec.~\ref{sec:InfoGeo}, the isKL embedding methods take the model manifold from the parameter space and position it an a hyperbolic using the symmetric Kullback-Liebler divergence. The KL-divergence functions similarly to a distance between models based on their (sufficient) statistics which determine the behavior of a particular model from the manifold. The isKL method thus embeds the model manifold in a behavioral space. This connection to the underlying behavior of the sampled models can be obscured when looking only at the results of the embedding analysis. As such, we will take some time here to discuss the behavior of the full-network model described by Eqn.~\ref{eqn:HawkesProc}.

\begin{figure*}[t]
\includegraphics[width=\linewidth]{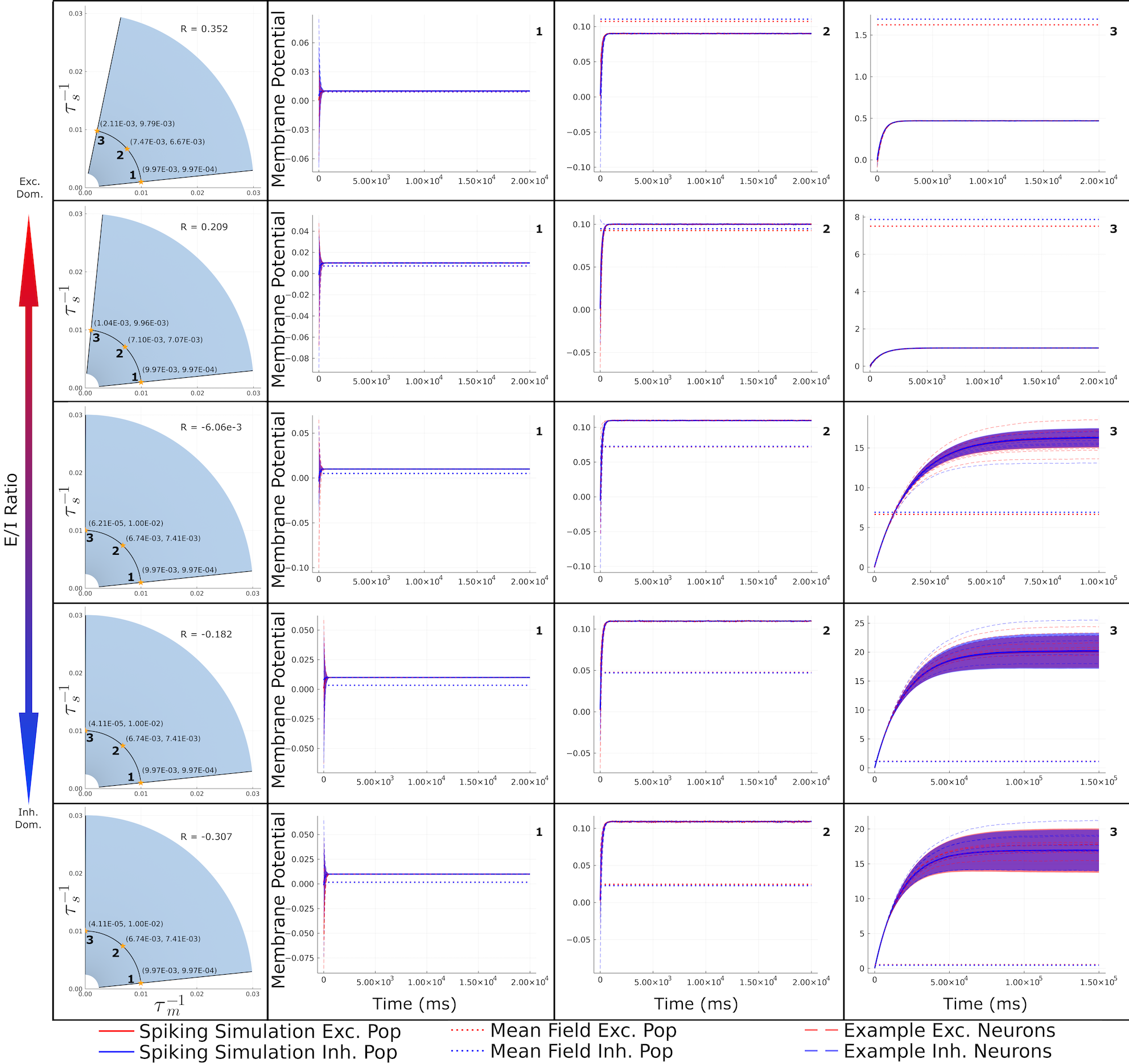}
\caption{\label{fig:LongTermMembrane} \textbf{Membrane potential dynamics of spiking network models} The dynamic population-averaged membrane potentials (solid lines) are plotted against the predicted mean-field values (dotted lines) for three different parameter pairs across the examined range of E/I ratios $R$. The membrane dynamics of six excitatory and six inhibitory neurons (dashed lines) are also shown for each condition. The ribbons around the population-averaged potentials are the the standard deviation of the membrane potentials within the corresponding population at each time point. The sampled parameter distribution from the embedding calculations, along with the chosen points for spiking simulation, are given in the right-most column. Full-network spiking simulations were run until an apparent stationary state was reached.}
\end{figure*}

Full-network spiking models were generated from the appropriate parameters in Table \ref{tbl:ModelParams} with a sparse, random connection matrix as described in Sec.~\ref{sec:NLHP}. A subset of model manifolds were chosen from across the range of E/I conditions $R$, and individual models from these manifolds were taken from along an arc of radius $\sim 0.01$ (see e.g.~Fig.~\ref{fig:LongTermMembrane}, left column). The membrane and spiking dynamics described by Eqn.~\ref{eqn:HawkesProc} with a specific choice of timescales were simulated using a basic forward-Euler integration scheme using a time step $dt=0.1$ ms. Most models were simulated for $20,000$ ms. The models in bottom 3 rows of the right-most column simulated for increased durations---$100,000$ ms, $150,000$ ms, and $150,000$ ms respectively---to assure convergence to a stationary behavioral regime. 

The long-term dynamics of the population-averaged membrane potentials and the membrane potentials of six excitatory and six inhibitory neurons are shown in Fig.~\ref{fig:LongTermMembrane}. The mean-field values of the membrane potentials predicted by Eqn.~\ref{eqn:Pop_MeanField} are represented by the red and blue dotted lines. From this figure, we see that the population-averaged membrane potentials in the full spiking network do indeed reach stationary values as predicted by the drift-eigenvalues $\lambda_A$ for the approximated spiking models shown in Sec.~\ref{sec:GPA_DriftEigs}. Additionally, we see that the mean-field membrane potential values correspond fairly well to the stationary population-averaged potentials (Fig.~\ref{fig:LongTermMembrane}, columns 1 and 2), but this breaks down near the upper bound of the arc (Fig.~\ref{fig:LongTermMembrane}, column 3). This upper boundary of the arc corresponds to the stability boundary in first two manifolds (Fig.~\ref{fig:LongTermMembrane}, rows 1 and 2) and a bifurcation boundary in the remaining manifolds. The breakdown of the mean-field approximations at these limits thus lines up with the colloquial understanding of their accuracy. Knowing that the population-averaged potentials of the full spiking networks converge to a stationary condition, we next look at how these stationary solutions of these models differ.

\begin{figure*}[t]
\includegraphics[width=\linewidth]{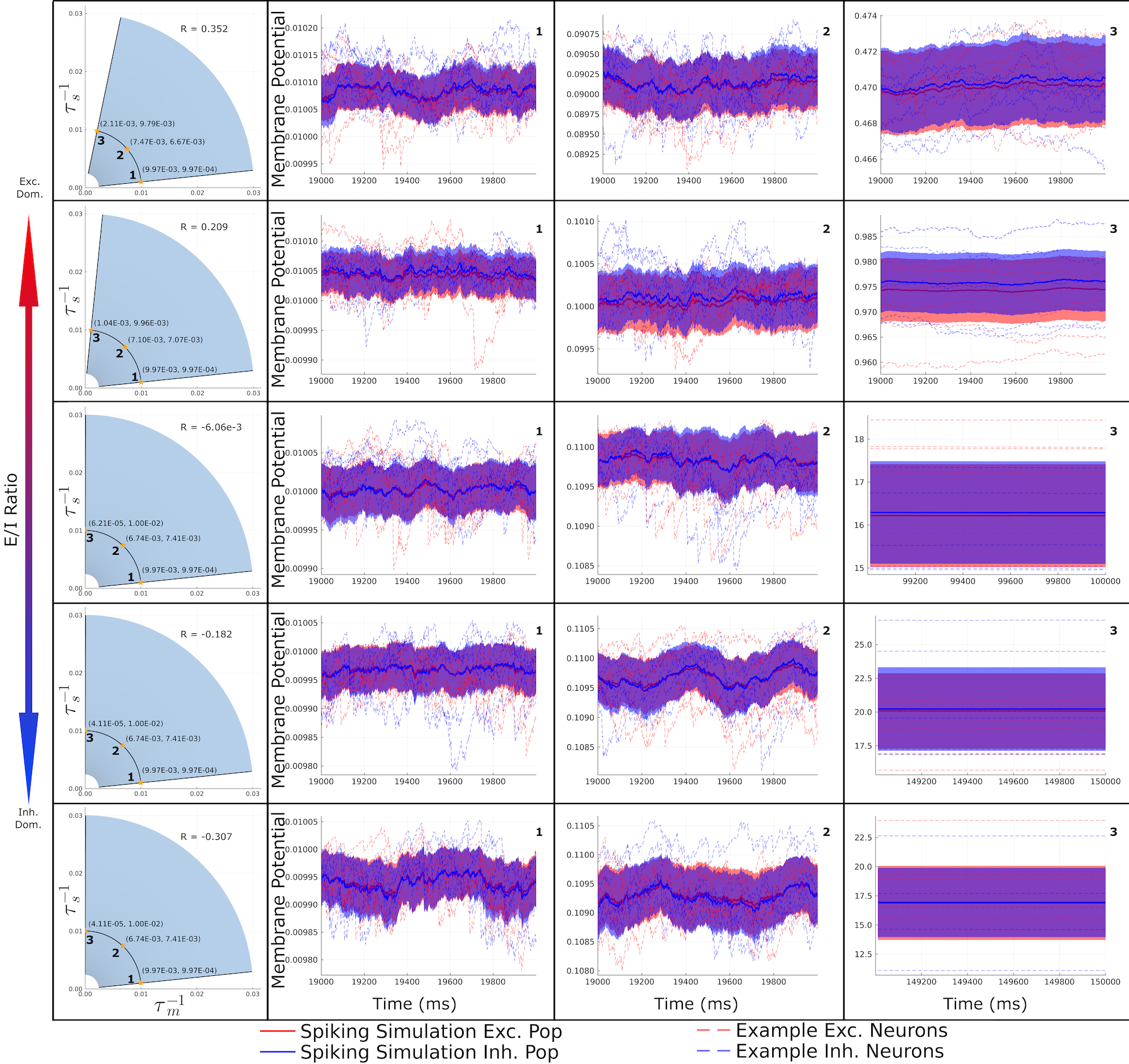}
\caption{\label{fig:StationaryMembrane} \textbf{Stationary membrane potential dynamics of spiking network models} The dynamic population-averaged membrane potentials (solid lines) are plotted against the predicted mean-field values (dotted lines) for three different parameter pairs across the examined range of E/I ratios $R$. The membrane dynamics of six excitatory and six inhibitory neurons (dashed lines) are also shown for each condition. The ribbons around the population-averaged potentials are the the standard deviation of the membrane potentials within the corresponding population at each time point. The sampled parameter distribution from the embedding calculations, along with the chosen points for spiking simulation, are given in the right-most column. Full-network spiking simulations were run until an apparent stationary state was reached, and the membrane dynamics for the the last $1000$ ms of simulation time are plotted.}
\end{figure*}

The membrane potential dynamics in the stationary regime of these same models are shown in Fig.~\ref{fig:StationaryMembrane}. Here, we show data from the last $1000$ ms of simulation and ignore the mean-field predictions. We see that the population-averaged membrane potentials visibly fluctuate around a average variable for most of the simulated models. These fluctuations in the population averages become less noticeable as the overall magnitude of the averages and standard deviations increase, e.g.~along column 3 of Fig.~\ref{fig:StationaryMembrane}. A similar trend is seen in the membrane dynamics for individual neurons in the network. Fluctuations in individual potentials are very large relative to their mean values for the first two models along the arc (Fig.~\ref{fig:StationaryMembrane}, columns 1 and 2). The population-variance in membrane potentials for these models is thus highly dependent on the fluctuations of individual membrane potentials. Towards the upper end of the arc (Fig.~\ref{fig:StationaryMembrane}, column 3), the magnitude of the membrane potentials increases and the fluctuations of individual potentials are less pronounced. For these models, the population-variance of the membrane potentials is much more dependent on the spread of individuals around the population-mean as opposed to the fluctuations of those individuals.

\begin{figure*}[t]
\includegraphics[width=\linewidth]{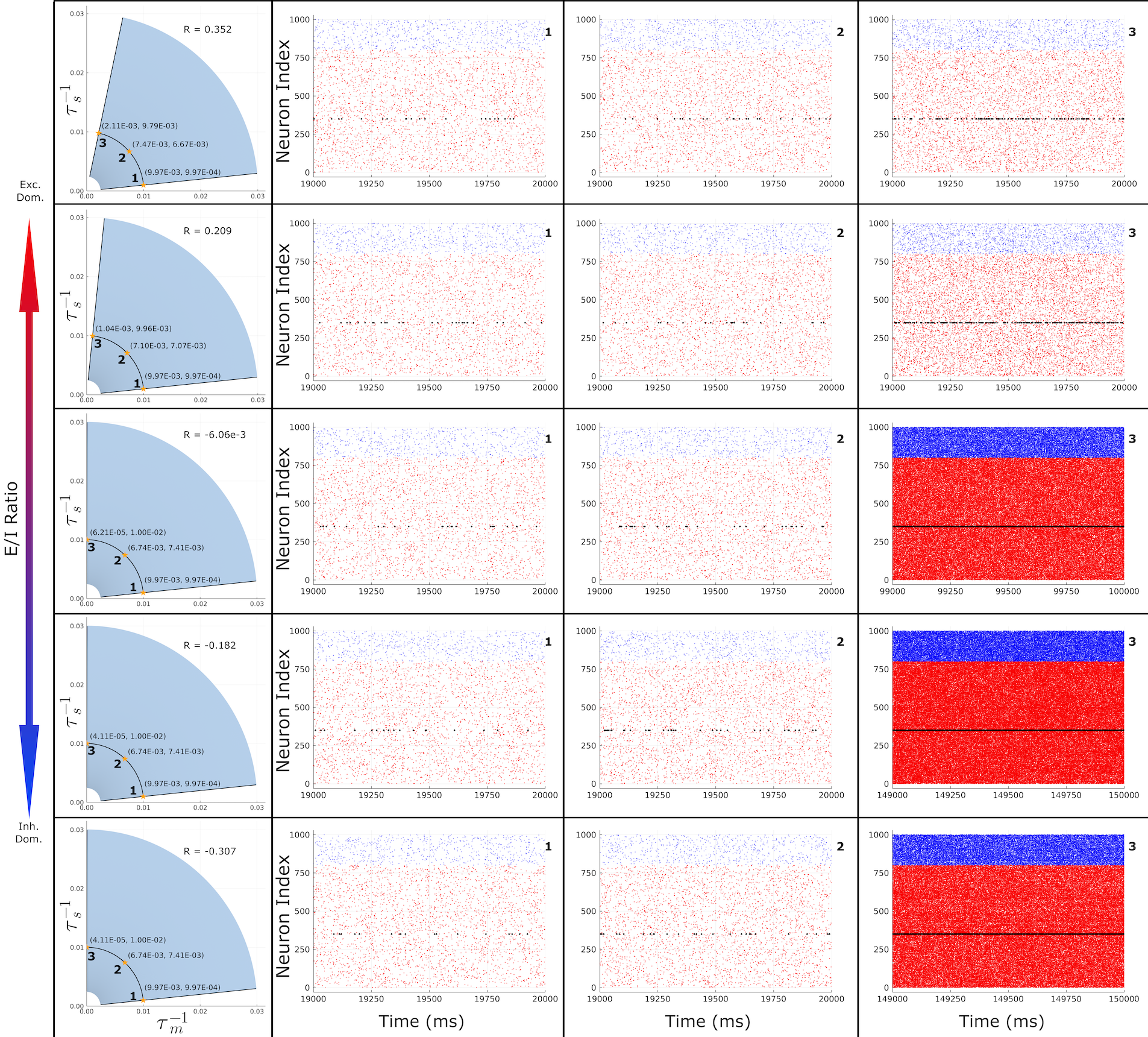}
\caption{\label{fig:StationarySpiking} \textbf{Stationary spiking behavior of full network models} Example raster plots for individual time-scale pairs for spiking models across the sampled range of E/I ratios $R$. The sampled parameter distribution from the embedding calculations, along with the chosen points for spiking simulation, are given in the right-most column. Full-network spiking simulations were run until an apparent stationary state was reached, and the spikes from the last $1000$ ms are plotted. In each model, neuron index $350$ was designated as the target neuron and its spikes are shown in black.}
\end{figure*}

We finish this section by examining the actual spiking dynamics for the example models in Fig.~\ref{fig:StationarySpiking}. Here we show raster plots for each model during the last $1000$ ms of simulation. First, we see that the target neuron (black spikes) fires more frequently than other excitatory neurons in the network, which is to be expected as it receives extra current input. For each model manifold (different rows in Fig.~\ref{fig:StationarySpiking}), we observe an overall increase in the rate of spiking in the network as we move along the arc from point 1 to point 3. This aligns with the change in overall magnitude of the membrane potentials seen in Figs.~\ref{fig:LongTermMembrane} \& \ref{fig:StationaryMembrane}. This observation also aligns with an intuitive understanding of the timescales: along this arc, the relative rate of synaptic input becomes much faster than the relaxation dynamics. This trend is taken to the extreme in the models at the top of the arc for each manifold (Fig.~\ref{fig:StationarySpiking}, column 3) where we see unrealistically high spiking rates in the last three rows. With this, we've built an intuitive understanding of the differences in behavior across the model manifolds and across E/I conditions. We now move on to the embedding analysis for these models.

\subsection{Network embedding is hierarchical}\label{sec:HierarchicalEmbedding}

It has been previously reported that biological models exhibit a hierarchy of sensitivities to different parameter combinations relative to some cost function on model behavior \cite{GutenkunstEtAl2007}. A similar hierarchical structure has been observed in the widths of model manifolds and the corresponding eigenvalues induced by a particular embedding, and a correlation between the widths and eigenvalues has also been noted \cite{TranstrumMachtaSethna2011,QuinnEtAl2019}. The current modeling and embedding differs from these prior cases in that we are embedding probabilistic models in behavior space. Considering also the limited dimensionality of the current embedding, it is unclear if this hierarchical property should manifest in the current system. We show below that the manifolds for models of the types in Eqns.~\ref{eqn:PopGPA} \& \ref{eqn:PopLFRN} are indeed hierarchical under the isKL embedding framework, with coordinate eigenvalues spanning several orders of magnitude for each E/I condition.

\begin{figure*}[t]
\includegraphics[width=\linewidth]{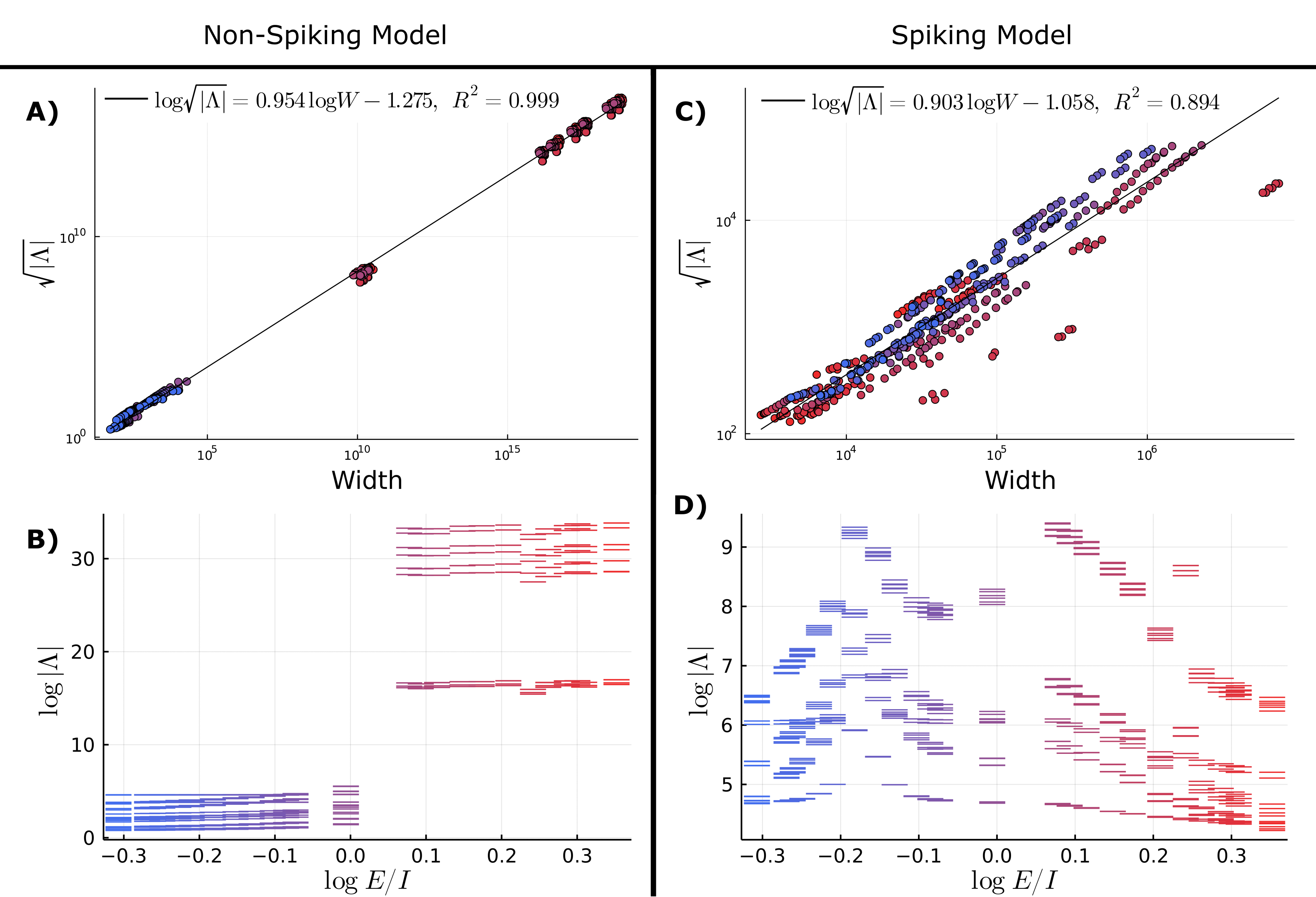}
\caption{\label{fig:EVdist} \textbf{Hierarchies of manifold widths} Top row: The correlation between the coordinate eigenvalues $\Lambda_i^{\pm}$ and the width $W$ across the manifold in that coordinate direction as the E/I ratio is varied for the non-spiking models (A) and spiking models (C). A simple linear regression is applied to the log-widths and log-absolute eigenvalues for visualization. The distribution of the scale of coordinate eigenvalues as the E/I ratio is varied for the non-spiking models (B) and the spiking models (D).  }
\end{figure*}

We used the isKL methods (Sec.\ref{sec:InfoGeo}) to embed the stationary distributions for both the spiking and non-spiking model types across 25 E/I conditions ranging from the excitation-dominated to the inhibition-dominated, and approximately centered at $R=0$. The root absolute eigenvalues for the embedding coordinates $\{\Lambda_i^\pm\}$ are plotted against the observed manifold width along the same coordinate for the non-spiking models (Fig.~\ref{fig:EVdist}A) and the spiking models (Fig.~\ref{fig:EVdist}C). Here, the manifold width is taken to be the simple range across a given coordinate. We see that the widths and eigenvalues are indeed correlated across E/I conditions for both model types, following with previous observations \cite{TranstrumMachtaSethna2011,QuinnEtAl2019}.This suggests these two measures may be used interchangeably in further analysis. The coordinate eigenvalues of the non-spiking model (Fig.~\ref{fig:EVdist}B) span at least three orders of magnitude for any given E/I condition tested, and up to nearly fifteen orders of magnitude at the most extreme. The coordinate eigenvalues for the spiking models (Fig.~\ref{fig:EVdist}D) span roughly two to three orders of magnitude on the extreme ends of the E/I spectrum and upwards of four in at some points in the center, with eigenvalues peaking towards the center as you approach from either extreme. Taken together, both model types studied here exhibit a hierarchical structure in line with prior observations of other systems, albeit with a more limited degree of separation in the case of the spiking-type models.

Before proceeding, we make some comparative observations between the two model categories. The scale and range of eigenvalues for the non-spiking model significantly larger than those for the spiking model type when in the excitation-dominated regime. Additionally, the non-spiking models show a sharp jump in eigenvalues when moving from the inhibition-dominated regime to the excitation-dominated one. This jump in eigenvalues may indicate a sort of bifurcation in the overall manifold. The eigenvalues $\{\Lambda_i^\pm\}$ directly reflect the covariance and---anecdotally more importantly---the variance in the corresponding sufficient statistics and natural parameters. A jump in the magnitude of the eigenvalues thus indicates a sudden increase in the variability of model behavior, and this could correspond to sampling near the stability boundary (Eqn.~\ref{eqn:StabBound}) in the case of the transition seen the non-spiking models. A similar transition may be happening at the peaks in the eigenvalue distributions of the spiking-type models, however these are much less pronounced than the one seen in the non-spiking models and the increase does not persist through the excitation-dominated regime as in the non-spiking models. We note that the firing rate non-linearity for the spiking model-type (Table \ref{tbl:ModelParams}) is quasi-linear when $x\gg 1$. A na\"{i}ve prediction would be a similar behavior between model types when the membrane potentials become more positive as in the excitation-dominated regime. However, this is not reflected in the observed distributions of $\{\Lambda_i^\pm\}$.

\subsection{Projection hierarchies}\label{subsec:ProjHier}
Having established that the isKL embedding of the spiking models and the non-spiking models exhibit a hierarchical structure, we next want to interrogate this structure across our model manifolds. We do this by examining projections of the manifolds onto lower-dimensional spaces along the largest widths and smallest widths. We will focus on $2$-dimensional projections. 

\begin{figure*}[t]
\includegraphics[width=0.7\linewidth]{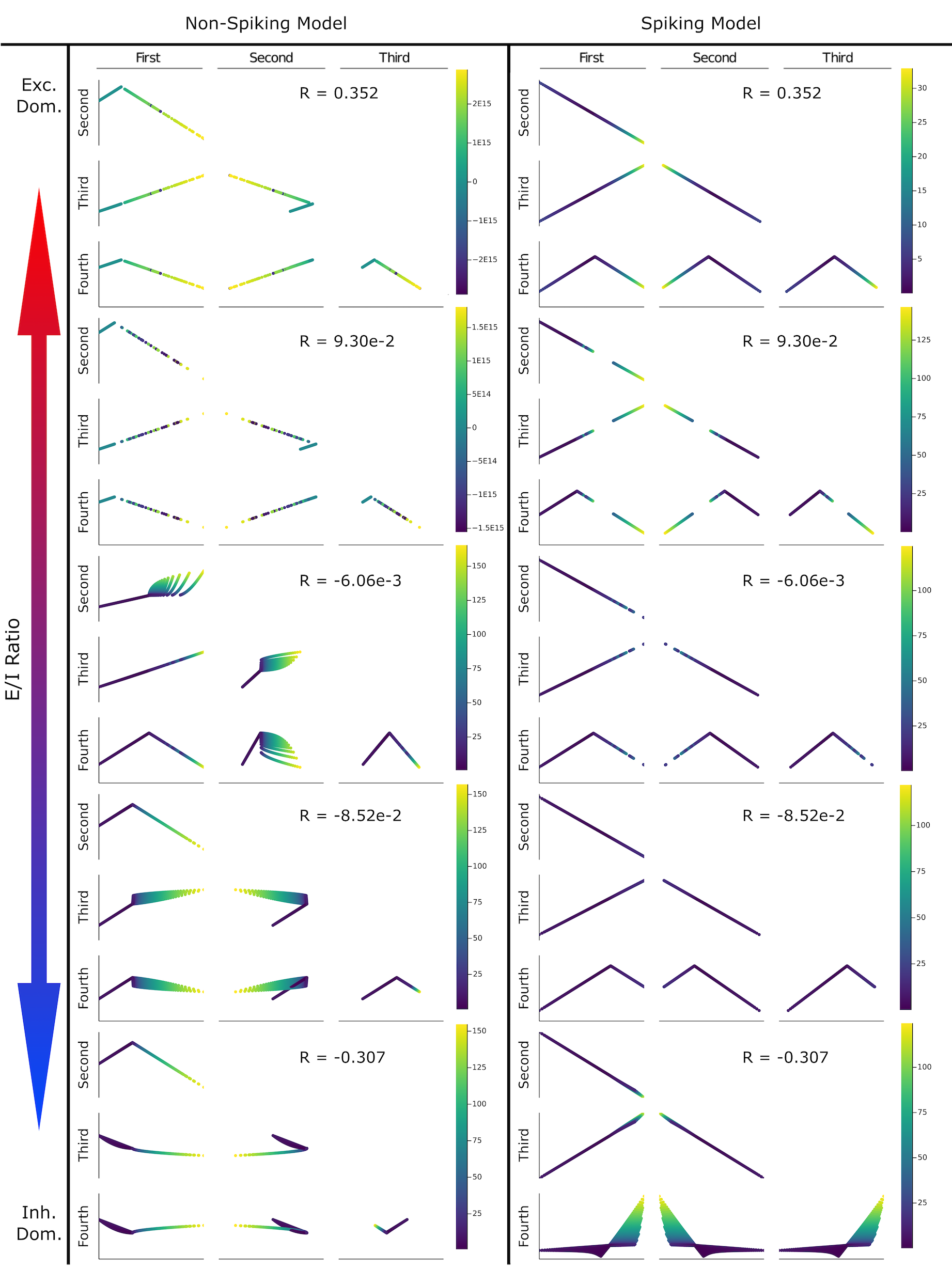}
\caption{\label{fig:ForwardHierarchy} \textbf{Largest manifold projections} Projections of the manifolds for the two model types onto the largest four coordinates as determined by the observed manifold width. The hierarchy of projections is shown as the excitation-inhibition ratio is changed from excitation dominant regime (top) to an inhibition-dominant regime (bottom). Manifolds are colored by $\langle V_0\rangle$---the mean membrane potential of the test neuron---and each projection is scaled by the largest observed width. These projections are the ``stiffest,'' contributing the most to the behavior of the distribution of activities.}
\end{figure*}

Fig.~\ref{fig:ForwardHierarchy} shows the largest manifold projections in behavioral space for the non-spiking models and spiking models across a subset of E/I conditions. Points on these manifolds are colored by the mean value of the membrane potential for the test neuron $\langle V_0\rangle $. It is visually clear that the manifold projections are shrinking from top left to bottom right for each condition, reflecting the hierarchical structure of the manifold. A large fraction of projections---for example, Fig.~\ref{fig:ForwardHierarchy} column 3, row 2---have apparent gaps in their structure. These are similar to the gaps seen in the projection of the example Gaussian distribution shown in Fig.~\ref{fig:CoordVis}G, and are tied to the sampling density used across the inverse-timescale space near key boundaries (data not shown).

Many of the projections across model types and E/I conditions appear very linear or piece-wise linear, for example Fig.~\ref{fig:ForwardHierarchy} column 2 rows 1-4. This thinness at the largest scales would suggest a relatively simple relationship between the largest coordinates and that the model manifold is relatively flat. The difficulty of overcoming under-sampling of the parameter space complicates this interpretation slightly. The gaps in the projections seen in the excitation-dominated regime are clear evidence of some under-sampling, but interpolating the data across gaps suggests that the projections in these conditions may still be piecewise- or quasi-linear. These stick-like projections both model-types in the excitation-dominated regime. The projections are also seen to change shape qualitatively as the E/I conditions change. In the non-spiking models, we see the appearance of spoon-shaped projections as we move into the inhibition dominated regime. In contrast we see knife-like projections in the spiking models, albeit only under the most inhibitory of E/I conditions. This qualitative change in the manifold projections seen across the two model types could be caused either by warping of the manifold along each coordinate as the conditions change or by changes in ranking of each coordinate. This point will be revisited in Sections \ref{subsec:CoordRanks} \& \ref{sec:SmoothChangingProjections}. 

Additionally, we note that many of the projections separate points on the manifold in alignment with $\langle V_0\rangle $, as was seen in the example embeddings shown in Fig.~\ref{fig:CoordVis}F \& G. This is particularly clear, for example, in the inhibition-dominated regime of the two model types (Fig.~\ref{fig:ForwardHierarchy} row 5). The separation of the manifold into sections based on behavioral regimes depends on more than just $\langle V_0\rangle $, however. For example, we see no such trend in Fig.~\ref{fig:ForwardHierarchy} column 2 row 4. In this case the individual models within the manifold are more significantly separated by (a combination of) behavioral parameters that, in a sense, have a dependence on the timescale parameters that is orthogonal to the way $\langle V_0\rangle $ depends on them. Exceptions aside, this noted $\langle V_0\rangle $-aligned separation serves as a clear demonstration of the behavioral clustering induced by the isKL method.

\begin{figure*}[t]
\includegraphics[width=0.7\linewidth]{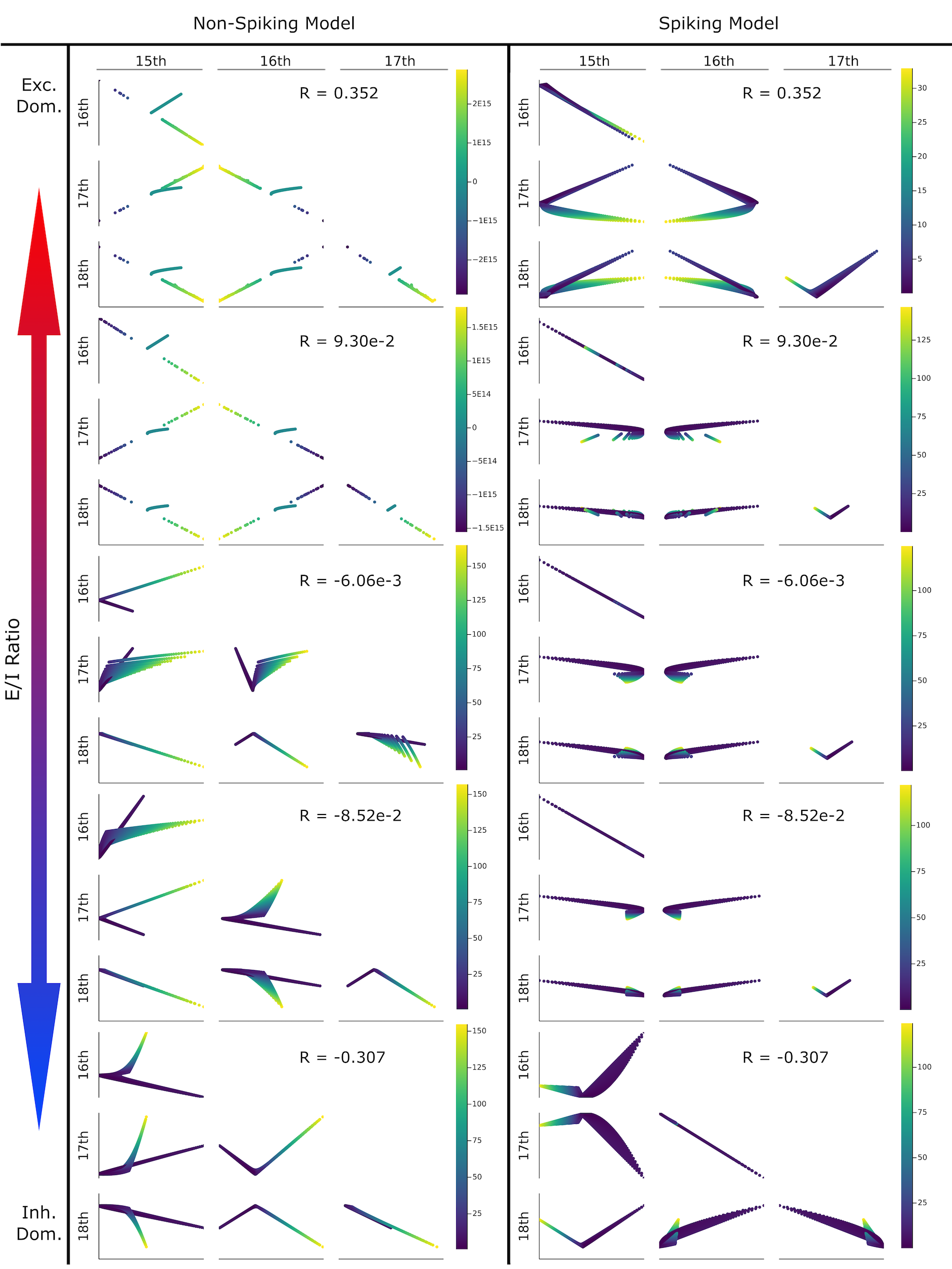}
\caption{\label{fig:ReverseHierarchy} \textbf{Smallest manifold projections} Projections of the model manifold for the two model types onto the smallest four coordinates as determined by the observed manifold width. The hierarchy of projections is shown as the excitation-inhibition ratio is changed from excitation dominant regime (top) to an inhibition-dominant regime (bottom). Manifolds are colored by $\langle V_0\rangle$, and each projection is scaled by the observed width along the coordinate ranked $(M-3)=15$. These projections are the ``sloppiest,'' contributing the least to the behavior of the distribution of activities.}
\end{figure*}

We can also examine the smallest projections of the model manifolds for the two model types, which correspond to the least important modes of the expansion in Eqn.~\ref{eqn:pdfexpansion}. Fig.~\ref{fig:ReverseHierarchy} shows the smallest projections of the model manifolds for all three model types across E/I conditions. As was the case for the largest coordinate projections, there is evidence of under-sampling in the smallest projections also. This particularly evident in the non-spiking model manifold in the excitation-dominated regime (Fig.~\ref{fig:ReverseHierarchy}, column 1, rows 1 and 2). In contrast to the largest projections, the stick-like projections comprise the minority of the small-projection shapes. The model manifolds appear instead to be highly curved at the fine-grained level. Following the observation from the large-scale projections, we see the smallest manifold projections separate points in line with $\langle V_0\rangle $. In particular, the counterexample mentioned before ( Fig.~\ref{fig:ForwardHierarchy} column 2 row 4) now also shows a degree of alignment with changes in $\langle V_0\rangle $, now in Fig.~\ref{fig:ReverseHierarchy} column 2 row 4. This highlights the fact that model separation along different directions on the manifold can be more or less tied to a particular behavioral or statistical parameter.

Before moving on, we should highlight the relationship between the projection coordinates and model behavior. Coordinates with larger eigenvalues contribute more to separation between $p(x|\theta)$ and $p(x|\theta')$ as measured by the $D_{sKL}$. More specifically, a relatively large absolute-eigenvalue $|\Lambda_i^\pm|$ indicates that the corresponding natural parameter is a relatively better way to separate individual models on the manifold by their behavioral predictions, or alternatively that a larger part of the variance of behavioral predictions across the manifold are explained by the associated natural parameter. Given a tractable mapping between the underlying model parameters and the natural parameters, the magnitude of coordinate eigenvalues can also give a sense of the relative importance of different parameter combinations aligning with a given coordinate direction. Lastly, the high degree of correlation between coordinate eigenvalues and the manifold widths along those coordinates means that the relative size of a particular projection gives a visual representation of the importance of the corresponding parameter combination.

\subsection{Coordinate rankings}\label{subsec:CoordRanks}

\begin{figure*}[t]
\includegraphics[width=0.8\linewidth]{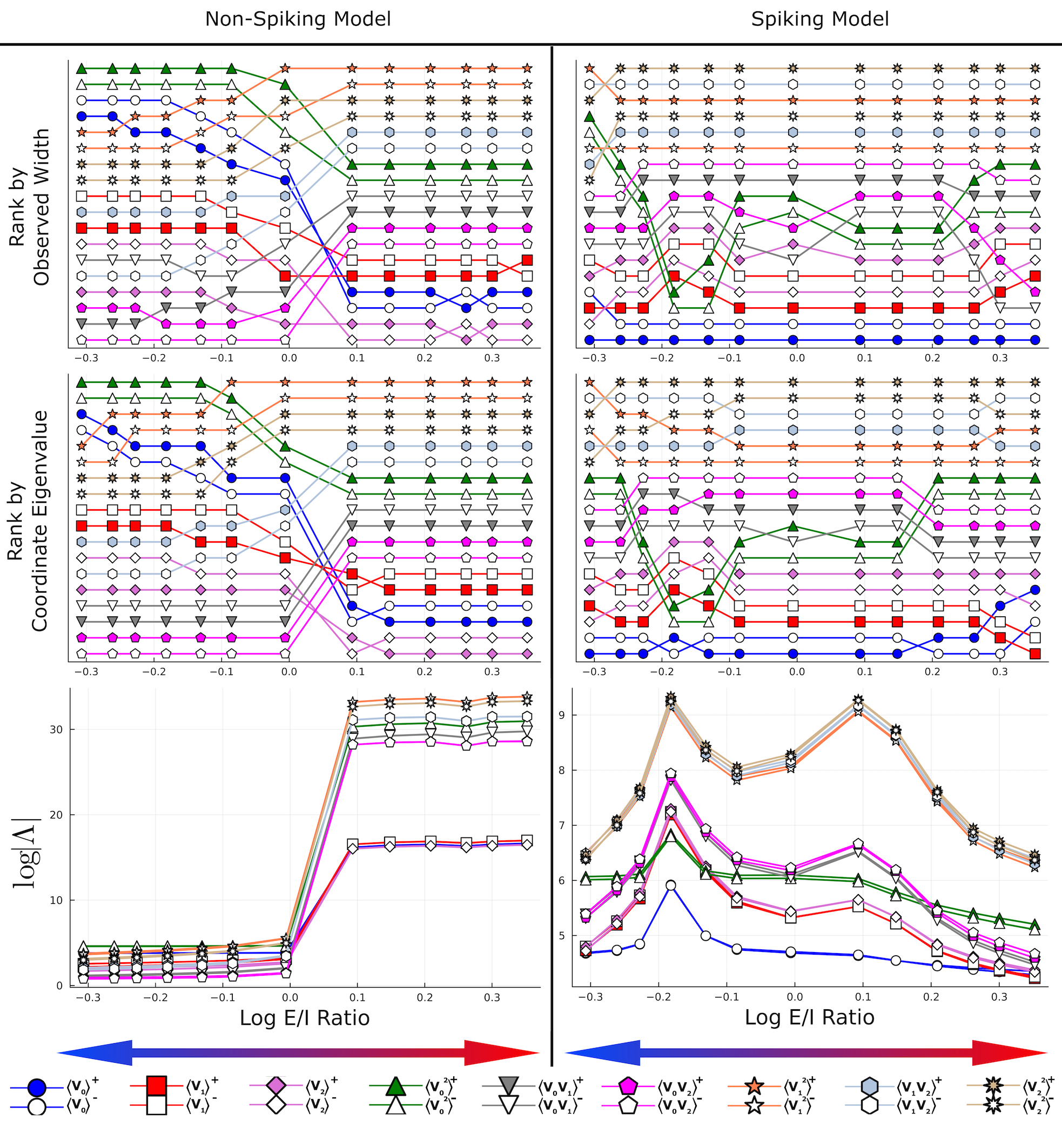}
\caption{\label{fig:CoordRanks} \textbf{Coordinate rankings} The ranking of each manifold coordinate $T_i^\pm$ as the E/I balance is changed in both model types. Coordinates are ranked from most important (top of each plot) to least important (bottom of each plot) based on the observed width of the manifold along said coordinate (top row) or the magnitude of the corresponding eigenvalue $|\Lambda_i^\pm|$ (middle row). The log-magnitude of the eigenvalue for each coordinate is given in the bottom plot as the E/I balance $R$ is changed as in Fig.~\ref{fig:EVdist}. The legend renames each coordinate $T_i^\pm$ to the corresponding sufficient statistic $\langle t_i\rangle^\pm$ for ready interpretation.}
\end{figure*}

We saw in Sec.~\ref{subsec:ProjHier} that the projections hierarchies of the model manifolds changed across the examined range of E/I conditions $R$. One possible explanation for this is that the rankings of coordinates by manifold width change with $R$. This potential aspect of the changing projections can be interrogated by tracking the rank of each coordinate across the range of $R$. This will additionally provide insight into what aspects of the statistical model have the greatest (or least) impact on the overall model behavior for both model-types. Fig.~\ref{fig:CoordRanks} depicts the ranking for each coordinate by both the length of the manifold along that coordinate (top row) and eigenvalue-magnitude (middle row) for the spiking and non-spiking model types for a subset of E/I conditions. As each coordinate corresponds directly to a single sufficient statistic, we color- and shape-code the rank of each coordinate according to this correspondence. The bottom row of Fig.~\ref{fig:CoordRanks} reproduces the eigenvalue distributions shown in Fig.~\ref{fig:EVdist}, except each point is now color- and shape-coded according to the sufficient statistic instead of the E/I measure $R$. 

We see in Fig.~\ref{fig:CoordRanks} that the ranking of coordinates changes across E/I conditions for both the non-spiking and the spiking models. We also note that the rankings by observed width (top row) and by eigenvalue-magnitude (middle row) agree fairly well across the range of $R$. This agreement between the two sets of rankings makes sense when considering the high degree of correlation between the coordinate eigenvalues and manifold widths shown in Fig.~\ref{fig:EVdist}. It is interesting that the eigenvalue distributions---particularly those for the spiking models---seem to separate into clusters of coordinates that do not intersect for much of the range of $R$. For example, the tan-orange-steel blue (eight point star-five point star-hexagon) cluster at the top of the eigenvalue spectrum corresponds to the second moments involving the inhibitory and excitatory populations. This cluster remains consistently above the grey-magenta (downward triangle-pentagon) and pink-red (diamond-square) clusters---which correspond to the second moments involving the target neuron and first moments for the bulk populations, respectively---across $R$ for the spiking models.

Knowing that the eigenvalues and thus the eigenvalue-magnitude rankings form these clusters across the E/I spectrum, it is natural to examine the sufficient statistics that correspond to coordinates in these clusters. We will focus on the spiking-type models that exhibit these clusters. The tan-orange-steel blue (eight point star-five point star-hexagon) cluster noted before dominates over other clusters in the spiking models, and these coordinates correspond to the second-order moments of the membrane potentials of the excitatory and inhibitory populations (Fig.~\ref{fig:CoordRanks} columns 2 and 3). This indicates that they are the most important statistics for distinguishing models across the manifolds. These second moments are also important for the non-spiking model manifolds, but they only sit at the top of the hierarchy in the excitation dominated regime $R>0$. The grey-magenta (downward triangle-pentagon) cluster in the spiking models corresponds to the mixed second moments involving the target neuron $V_0$ and either $V_1$ or $V_0$. This cluster is above the green (triangles) cluster in the mid-range of $R$ and just below it in the extreme E/I conditions, and this green (triangles) cluster is the pure second moment $\langle V_0\rangle$. The green cluster is generally above the pink-red (diamonds-squares) and the blue (circles) clusters, except for a brief crossing of the pink-red and the green clusters around $R\approx-0.18$. These last two clusters correspond to the mean values of all three membrane potentials. Taken together, these observations say that for the spiking-type models the fluctuations are more important for distinguishing between individual models on a given manifold and---for both the first and second moments---the statistics that involve the test neuron are generally less important than those that do not. These observations hold for the non-spiking model manifolds in the excitation-dominated regime, but not in the inhibition-dominated regime (Fig.~\ref{fig:CoordRanks}, column 1).

Let us discuss the coordinate rankings at a more granular level of detail. While both model types have the second moments at the top of their respective hierarchies in the excitation-dominated regime, it is interesting to note how they differ here. The spiking-type models have the $\langle V_2^2\rangle^+$-related coordinates at the top while the non-spiking model is topped by the $\langle V_1^2\rangle^+$-related coordinates. The suggests that the degree of fluctuations in the inhibitory population are the most varied for the spiking models in this regime, but the excitatory population fluctuations take that title in the excitation-dominated non-spiking models. The last fine-grained detail we highlight here is the increased importance of the $V_0$-moments in distinguishing the behavior of the inhibition-dominated non-spiking models relative to their importance in the inhibition-dominated spiking-type models.

In addition to visualizing the relative importance of certain sufficient statistics for distinguishing between particular models across a given model manifold, we get another piece of information visualized for free through the eigenvalue-magnitude ranking plots (Fig.~\ref{fig:CoordRanks}, middle row). Recall from Eqn.~\ref{eqn:coordEVs} that the eigenvalues $\Lambda_i^\pm$ are given by the covariance of the sufficient statistic and natural parameter across the manifold (${\rm Cov}(\eta_i,\langle t_i\rangle)$) and the geometric means of their individual variances ($\sqrt{\var(\eta_i)\var(\langle t_i\rangle)}$). As we know the $\Lambda_i^-$ eigenvalues are negative and of the same order of magnitude as the corresponding $\Lambda_i^+$ (Fig.~\ref{fig:CoordRanks}, row 3), we know that geometric mean of those variances greatly outweighs their covariance. Furthermore, the relative ranking of $\Lambda_i^+$ and $\Lambda_i^-$ imply the sign of the covariance ${\rm Cov}(\eta_i,\langle t_i\rangle)$: if $\Lambda_i^+>\Lambda_i^-$ then ${\rm Cov}(\eta_i,\langle t_i\rangle)>0$ and \emph{vice versa}. For example, by looking at the eigenvalue-magnitude ranking of the steel blue (hexagon) coordinates in the spiking-type models (Fig.~\ref{fig:CoordRanks}, row 2 columns 2 and 3) we see that ${\rm Cov}(-\frac{1}{2}C_{12}^{-1},\langle V_1V_2\rangle)<0$ across all E/I conditions examined here. While this could very easily be determined by looking at these covariances themselves---and they must be calculated in order to determine $\Lambda_i^\pm$---it is convenient to be able to glean this from a plot already produced for another purpose.

We will briefly summarize. Fig.~\ref{fig:CoordRanks} shows that the coordinate rankings do change across the sampled E/I range, thus explaining the changing projection hierarchies in Figs.~\ref{fig:ForwardHierarchy},\ref{fig:ReverseHierarchy} at least in part. We found that the coordinates form clusters in the eigenvalue distribution that behave in a correlated manner across the E/I spectrum and with which they share relations to similar types sufficient statistics. In particular, the cluster of coordinates corresponding to the fluctuations $\langle V_IV_J\rangle^\pm$ for $I,J\in\{1,2\}$ have the most impact on the activity of the spiking-type models, as well as in the excitation-dominated non-spiking models. We made observations of which types of fluctuations were most important to model distinction across the manifold for different model-types and different E/I conditions. Finally, we highlighted a secondary visual interpretation of the eigenvalue-magnitude ranking plots relating to the base statistical model.

\subsection{Transforming of base parameters}\label{sec:ParamTransforms}

\begin{figure}[t]
\includegraphics[width=\linewidth]{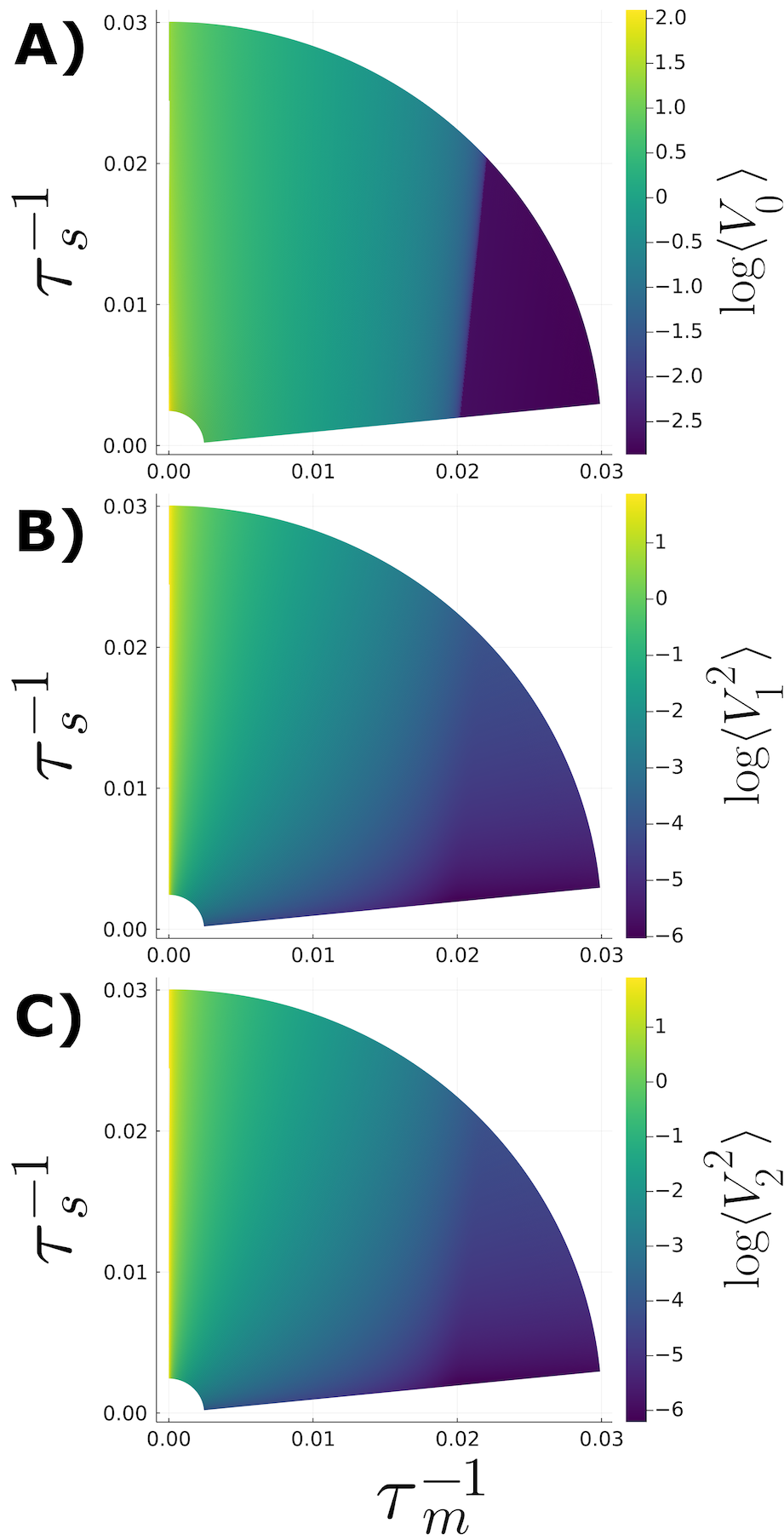}
\caption{\label{fig:ParamTransforms} \textbf{Mapping from inverse timescales to statistical parameters} The relationship between the inverse timescales $(\tm^{-1},\ts^{-1})$ and a select set of statistical parameters from the corresponding stationary Gaussian distribution for the spiking network with $R=-6.06\times 10^{-3}$. Inverse timescale spaces are colored by the log-value of one of the sufficient statistics: A) $\log_{10}\langle V_0 \rangle$; B) $\log_{10}\langle V_1^2 \rangle$; C) $\log_{10}\langle V_2^2 \rangle$} 
\end{figure}

We highlighted in Sec.~\ref{subsec:CoordRanks} that the statistical parameters from the stationary Gaussian distribution of membrane potentials---discussed in terms of the sufficient statistics---have a hierarchical impact on the possible behaviors of the spiking model that changes across the E/I spectrum $R$. Further, we identified clusters of parameters that tended change in similar ways with $R$. In the case of the spiking models, the fluctuations $\langle V_IV_J\rangle^\pm$ for $I,J\in\{1,2\}$ were the most impactful while the mean coordinates $\langle V_0\rangle^\pm$ had a relatively small impact. While important, these observations do not directly address the role of the inverse timescales $(\tm^{-1},\ts^{-1})$ on model behavior. Unfortunately, the mapping from the timescale parameters to statistical parameters is intractable, owing primarily to the transcendental system of mean-field equations (Eqn.~\ref{eqn:Pop_MeanField}). Closed forms for the stationary distribution parameters of the linear non-spiking models can be found, but these expressions are ratios of high-degree polynomial functions of the timescales and do not directly reflect the mapping in the spiking model context. To begin untangling the impact of the timescale parameters on the range of model behaviors, we must thus rely on a qualitative understanding of the relationship between the timescales and e.g.~the sufficient statistics.

In Fig.~\ref{fig:ParamTransforms}, we plot the logarithm of several sufficient statistics as a function of position in inverse-timescale space for the spiking network with $R=-6.06\times 10^{-3}$. We include $\langle V_1^2\rangle$ (Fig.~\ref{fig:ParamTransforms},B) and $\langle V_2^2\rangle$ (Fig.~\ref{fig:ParamTransforms},C) from the upper cluster as well $\langle V_0\rangle$ (Fig.~\ref{fig:ParamTransforms},A) to serve as a representative set from across the hierarchies in Fig.~\ref{fig:CoordRanks}. Note that the second moments have different units than those of $\langle V_0\rangle$, which should be kept in mind when comparing the color scales. That said, the $D_{sKL}$ between two members of the same exponential family can be rewritten as \cite{TeohEtAl2020}
\begin{align*}
    D_{sKL}(\theta,\theta')=&\sum_i \left(\eta_i(\theta)-\eta_i(\theta') \right) \left(\langle t_i(\theta)\rangle-\langle t_i(\theta')\rangle \right).
\end{align*}
Paired with the sufficient statistics of a multivariate normal distribution (Eqn.~\ref{eqn:natparams_suffstats}), we see that the $D_{sKL}$ is in some sense weighing the first and second moments directly against each other. This in mind, the variability in the second moments is $\sim 2$ orders of magnitude larger than that for the mean of the test neuron, in line with their relative ranking in Fig.~\ref{fig:CoordRanks}. Further, we note the similar dependence of all three statistical parameters on the inverse timescales, increasing in magnitude radially from $\tm^{-1}$-axis to the $\ts^{-1}$-axis as well as exhibiting a ``cold spot'' triangle on the right-most corner of the sampled wedge. The trends between the means and covariances differ most significantly along the $\ts^{-1}$ boundary. Here, the magnitude of the mean increases towards the inverse-timescale origin (i.e.~very long timescales) while the second moments increase moving away from the origin (i.e.~very short timescales).

The presence of the cold spot in the mappings to the statistical parameters---particularly the sharpness of the transition seen for $\langle V_0\rangle^\pm$---reinforce the intuition that translating changes in the statistical parameters back onto the timescale parameters is non-trivial. That said, the shared general trend in the mappings suggest a possible avenue for model reduction if some loss of expressivity is permitted. Reducing the 2-dimensional sampled space to an arc around the origin and through the cold spot could be used to capture the concomitant increases in the magnitude of the first and second moments, capturing the majority of their respective variability. Alternatively, radial sampling along the $\ts^{-1}$ boundary could be used to study the apparent trade-off in magnitude of the means and covariances. This idea of model reduction is intimately tied to notions of model dimensionality, which we will return to in Sec.~\ref{sec:Dimensionality} and Sec.~\ref{sec:Discussion}.

\subsection{Manifold projections change smoothly with E/I balance}\label{sec:SmoothChangingProjections}

\begin{figure*}[t]
\includegraphics[width=0.89\linewidth]{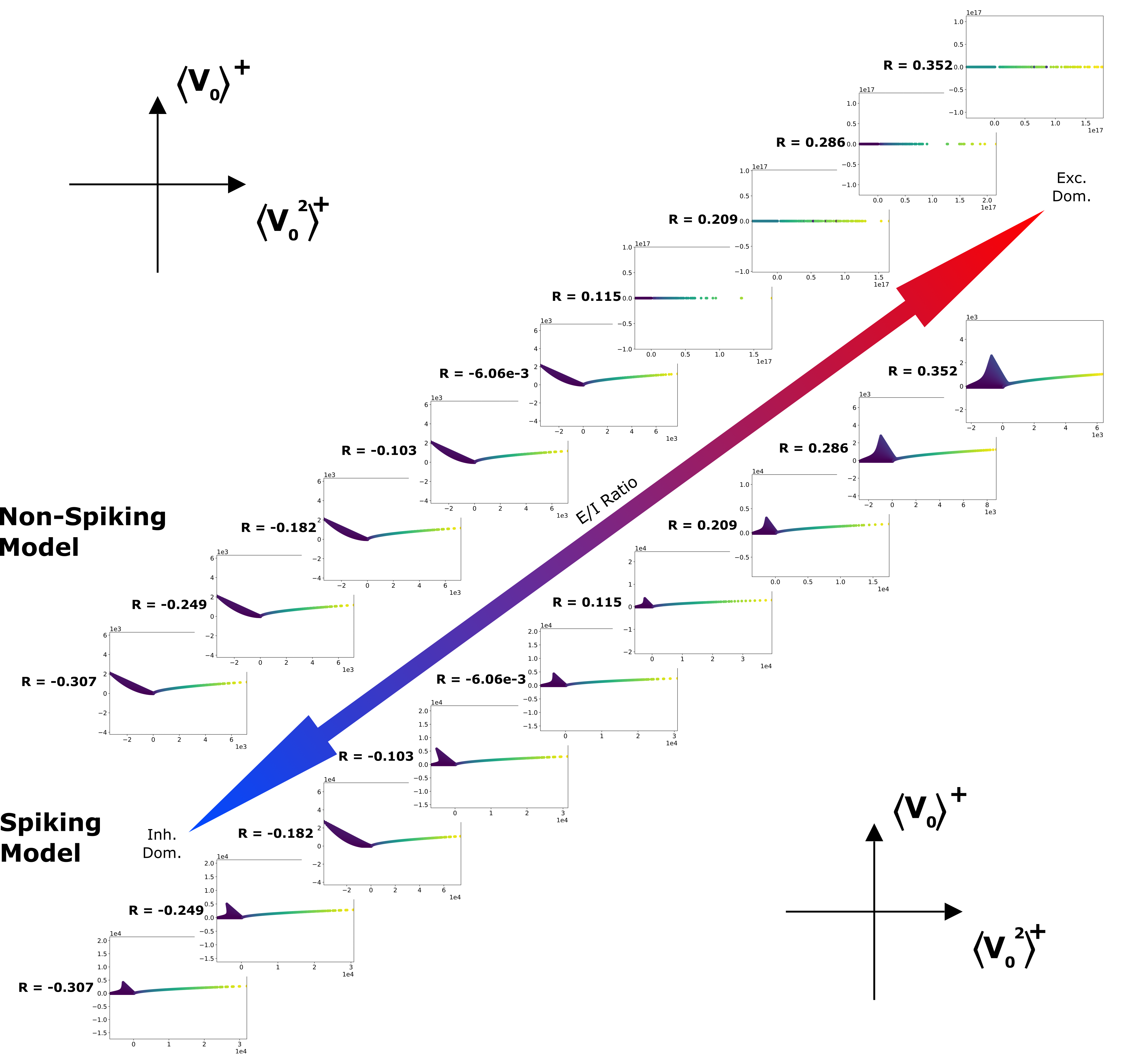}
\caption{\label{fig:CoordEvol} \textbf{Coordinate evolution as E/I balance is tuned} Projections onto the $\langle V_0^2\rangle^+$-$\langle V_0\rangle^+$ plane of the non-spiking model manifolds (top) and the spiking model manifolds (bottom) as the excitation-inhibition ratio is adjusted. The axes of each 2-D projection are scaled to the larger width for the given model and E/I condition for the convenience of visualization. Note that the projections change in size by an order of magnitude as the E/I balance is adjusted. The projections are colored by $\langle V_0\rangle$ as in prior figures.} 
\end{figure*}

We now return to the question raised at the beginning of Sec.~\ref{subsec:CoordRanks}: What causes the projection hierarchies to change across E/I conditions? While the changing coordinate ranking observed across E/I conditions for both models can explain the changing manifold projections, it does not rule out the possibility that the projections along a given coordinate are themselves changing. To address this possibility, we project the model manifolds for both of the model-types onto the same pair of coordinates across the E/I spectrum in Fig.~\ref{fig:CoordEvol}. We chose to project the manifold onto the space-like $\langle V_0^2\rangle^+$ and $\langle V_0\rangle^+$ coordinates as the statistical behavior of the test neuron may be of particular interest in some scenarios. For the sake of visualization, each projection along the E/I spectrum is scaled by the larger of the two manifold projections at each condition. The overall scale of the projection is given by the axis scale. 

We can see in the projections of the non-spiking model manifolds (Fig.~\ref{fig:CoordEvol}, upper diagonal) that there is a squashing and stretching of the manifold relative to the overall change in scaling as the E/I conditions are changed. Additionally, these transformations appear to act smoothly on the manifold projections until the manifold flattens going from the inhibition-dominated regime to the excitation-dominated one in the range $-6.06\times 10^{-3}\leq R\leq 0.115$. This flattening reflects a radical increase in the manifold scale along the $\langle V_0^2\rangle^+$-coordinate relative to the $\langle V_0\rangle^+$-coordinate as all of the eigenvalues are seen to jump (Fig.~\ref{fig:CoordRanks}, column 1 row 3). This interpretation is corroborated by the change in overall scale of the axes---from $\sim\BigO(10^3)$ for $R<0.115$ to $\sim\BigO(10^{17})$ for $R\geq0.115$ (Fig.~\ref{fig:CoordEvol}, upper diagonal)---and the correlation between eigenvalue-magnitude and manifold width discussed in Sec.~\ref{sec:HierarchicalEmbedding} (Fig.~\ref{fig:EVdist}). The projections of the spiking manifolds (Fig.~\ref{fig:CoordEvol}, lower diagonal) are also seen to transform smoothly with $R$ with the fork-shaped projections (e.g.~lower diagonal, $R=-0.307$) collapsing into the spoon projections (e.g.~lower diagonal, $R=-0.103$) seen in the non-spiking model around $R\approx -0.1$. The projections for the spiking-type manifolds do change along $R$ in line with the changes in their respective eigenvalue distributions (Fig.~\ref{fig:CoordRanks}, columns 2 and 3, row 3), but these changes are more subtle than in the non-spiking model. The eigenvalue distributions for the spiking-type models drift downwards as you move from $R<-0.18$ to $R>0.10$, and this is mirrored in the manifold projects by a slight decrease in projection scale moving in the same direction.

We have shown here that the manifold projections for both model-types do indeed change across the sampled E/I range, which plays a subsequent role in the changing of projection hierarchies across E/I conditions noted in Sec.~\ref{subsec:ProjHier}. The individual projections were shown to undergo potentially significant rescaling across values of $R$ that alter it visually, as noted in the non-spiking manifold. Additionally, the manifold can exhibit a warping, as in the fork-spoon-fork transition noted in the spiking-type models.

\subsection{Manifold dimensionality}\label{sec:Dimensionality}

\begin{figure}[t]
\includegraphics[width=\linewidth]{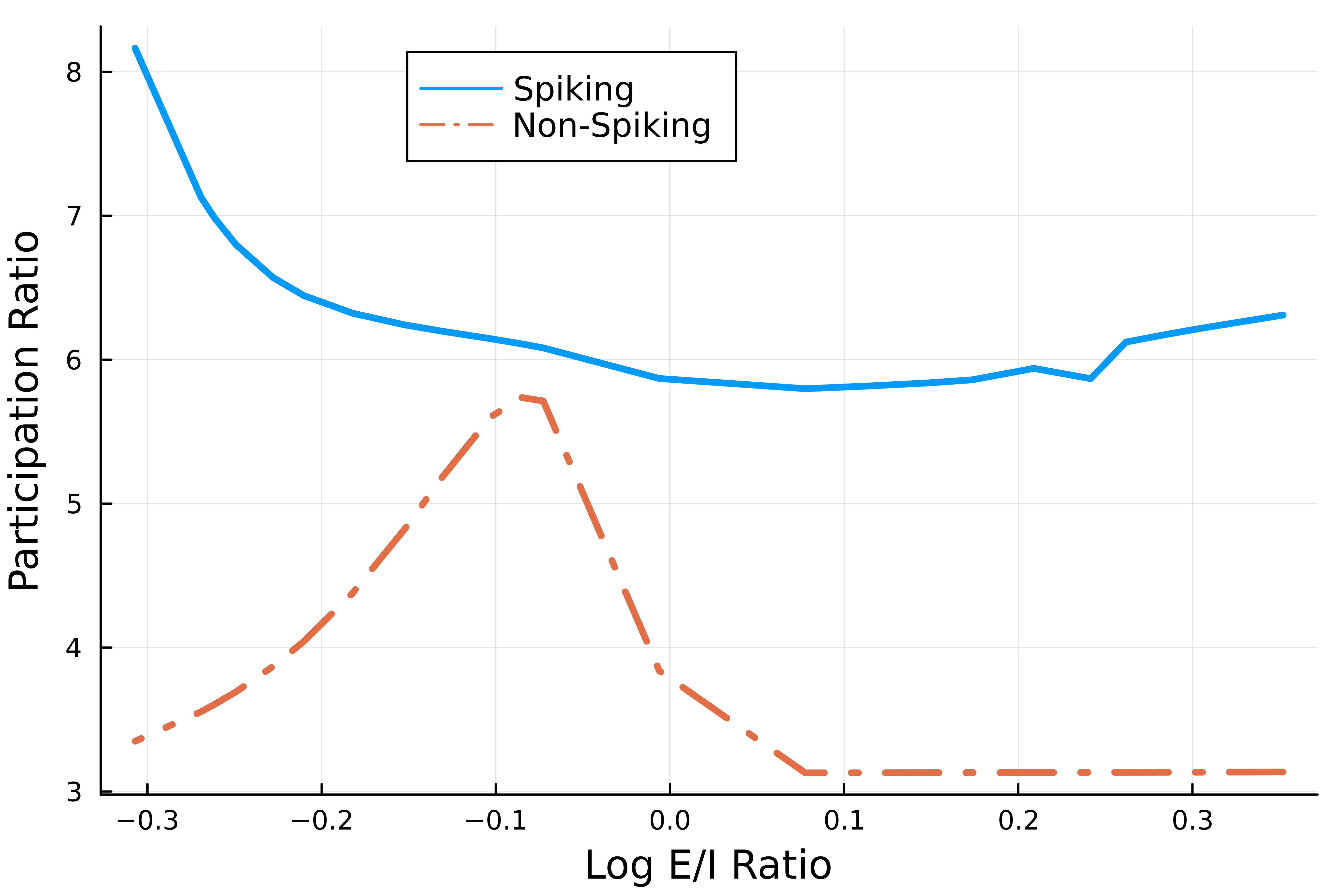}
\caption{\label{fig:ManifoldDimensionality} \textbf{Dimensionality of model manifolds} The effective dimensionality of non-spiking and spiking models across the spectrum of E/I conditions is given. Effective dimensionality is given by the altered participation ratio given by Eqn.~\ref{eqn:AltPR}.} 
\end{figure}

As mentioned at the outset (Sec.~\ref{sec:Intro}), a key issue when analyzing the behavior of a collection of large spiking networks is the dimensionality of the behavioral output space. The true behavioral space of the model---as expressed through the spiking activity---grows with both an increasing network size $N$ and a decreasing time bin size $\Delta t$. A goal of the current work is to understand the behavioral output of these models in a lower dimensional framing. Thus, we will briefly interrogate the dimensionality of our spiking and non-spiking models before proceeding to the final discussion.

The behavioral dimensionality of models following Eqn.~\ref{eqn:HawkesProc} is $NT/\Delta t$ for a discrete-time trial of length $T$. The dimensionality of the behavioral space remains the same when moving to the Gaussian process approximation of the model in Eqn.~\ref{eqn:NetwGPA}. Population-averaging of these approximated dynamics into Eqn.~\ref{eqn:PopGPA} decreases the dimensionality to $N_{\rm pop}T/\Delta t$, where $N_{\rm pop}<N$ is the number of populations being considered. By simplifying our analysis to studying the \emph{stationary distribution} of population behaviors, the behavioral dimensionality drops to $N_{\rm pop}(N_{\rm pop}+3)/2$ corresponding to the maximal number of independent sufficient statistics (see Sec.~\ref{sec:InfoGeo}). Finally, the isKL methods embed this manifold of behaviors into an $N_{\rm pop}(N_{\rm pop}+3)$-dimensional space which determines the upper limit of dimensionality that may be measured from embedded data (i.e.~sampled models).

Having considered how the dimensionality changes across the steps for our analysis, two key questions remain. If we can only see the results of the embedding, how do we gauge the dimensionality of the manifold being analyzed? If we instead have an understanding of the maximal dimensionality of the system, is there any effective reduction in dimensionality that we can measure? To adress these questions, we start with a measure of effective manifold dimensionality commonly used in principal component analysis (PCA) known as the participation ratio (PR):
\begin{equation*}
    PR = \frac{\left(\sum_i\Lambda_i\right)^2}{\sum_i \Lambda_i^2}.
\end{equation*}
As the isKL embedding methods are intimately tied to multidimensional scaling (MDS)---an extension of PCA---the PR should serve as a useful base for measuring the effective dimensionality of our embedded model manifolds. This is complicated slightly by the presence of negative eigenvalues $\{\Lambda_i^-\}$ that arise in MDS, so we use an altered PR as our measure of effective dimensionality:
\begin{equation}
    PR = \frac{\left(\sum_{i,\pm}|\Lambda_i^\pm|\right)^2}{\sum_{i,\pm} \left(\Lambda_i^\pm\right)^2}.\label{eqn:AltPR}
\end{equation}

The effective dimensionality of our two model-types across the examined E/I spectrum is shown in Fig.~\ref{fig:ManifoldDimensionality}. We see that the spiking-type models begin with a relatively high $PR\approx 8$ in the inhibition-dominated regime before dropping to $PR\approx 6$ in the middle regime and then rising slightly again in the excitation-dominated regime. By contrast, the non-spiking model has $PR\approx3$ in the excitation-dominated regime. The $PR$ for the non-spiking model then peak at $PR\approx6$ around $R\approx-0.1$ before decaying back down further into the inhibition-dominated regime. The non-spiking models thus have a lower effective dimensionality than the spiking-type models.

How do we contextualize the measured $PR$ for these models? First, we note that the maximal possible measured dimensionality for both the spiking and non-spiking model-types is $N_{\rm pop}(N_{\rm pop}+3)=18$, and the \emph{statistical model} dimensionality is $N_{\rm pop}(N_{\rm pop}+3)/2=9$. This indicates that the approximate models show a dimensionality reduction compared to both the model dimensionality and the maximal embedding dimensionality. This seems trivial until one examines the $PR$ measure for the example embeddings given in Sec.~\ref{sec:InfoGeo}. The simple Poisson example has just 1 parameter $\Lambda$ and subsequently a maximum embedding dimension of 2. Despite the intrinsic parameter density of 1, its isKL embedding (Fig.~\ref{fig:CoordVis}F) gives an effective dimensionality is much closer to its maximal embedding dimension and gives $PR\approx1.982$. In a similar vein, the example Gaussian model has two parameters $(\mu,\sigma)$, yet it has an embedding dimension of $PR\approx3.929$ which is nearly its maximum embedding dimensionality of 4 (see Fig.~\ref{fig:CoordVis}G for one of the manifolds projections). The measured $PR$ thus does not seem to reflect the dimensionality of the intrinsic manifold structure, but instead the number of embedding dimensions within the isKL framework required to capture most of the variability in model behaviors. This will be discussed further in Sec.~\ref{sec:Discussion}.

\section{Discussion}\label{sec:Discussion}

The central motivation of this paper is to tease apart the impact of cellular and synaptic model parameters---internal timescales and relative synaptic strengths, respectively---on the complex and high-dimensional behavioral space of spiking network models. 
Taking inspiration from prior applications of information geometry to neural systems \cite{NakaharaAmari2002,WuAmariNakahara2002,AmariEtAl2003,AmariParkOzeki2006,ShimazakiEtAl2012,AmariKarakidaOizumi2019,AmariKarakidaOizumi2019-2,KarakidaAkahoAmari2020}, we approached this Herculean task by leveraging recently developed methods for studying the information geometry of complex biology models \cite{QuinnEtAl2019,TeohEtAl2020} and applying them to spiking network models with more biological features than those considered previously. We began by defining our spiking model \cite{OckerEtAl2017,BrinkmanEtAl2018} and then simplifying it through population-averaging, using a path-integral formalism to approximate the membrane dynamics as a Gaussian process \cite{OckerEtAl2017}, and then calculating the stationary distribution for that approximation \cite{VatiwutipongPhewchean2019}. The stationary distributions for these were then analyzed using the information geometric framework introduced by Teoh and colleagues \cite{TeohEtAl2020}. This workflow is the core of the work presented.

Before diving into the results of the geometric embedding analysis, we briefly examined the behaviors of full spiking networks across various E/I conditions and for a few different timescale points. We showed that the behavior of the actual networks change distinctly across the variables at both the level of spiking and of population-averaged membrane dynamics. Importantly, the spiking models reach a stationary behavior in the long-time limit. This agreed qualitatively with the mean field predictions and supported the analysis of the stationary distributions from the reduced model.

The information-geometric analysis demonstrated that the approximated models are hierarchical. Manifold widths and coordinate eigenvalues spanned several orders of magnitudes, pointing to a ``hyperribon'' structures with ``stiff'' and ``sloppy'' coordinate directions. The distribution of these coordinate eigenvalues changed across E/I conditions and with it the hierarchy of 2-dimensional manifold projections. These changes in the manifold projections arose from a smooth warping of projections onto specific coordinate pairs as well as a reordering of the coordinate rankings. Identifying each coordinates with their corresponding sufficient statistic highlighted a clustered structure in the eigenvalue distribution of the spiking models across E/I conditions. From this clustered structure, it is possible to pick out the most and least important sufficient statistics for distinguishing between models on a given manifold---these are the parameter combinations that underlie the stiff and sloppy coordinate directions, respectively. In particular, the stiffest directions on the spiking-model manifolds corresponded to the second moments of the excitatory and inhibitory population membrane potentials while the sloppiest directions were those corresponding to the first moments. This suggests that bulk fluctuations are key for determining the behavior of a specific network. It is unfortunately difficult to tie this understanding of stiff and sloppy statistical parameters to the timescale parameters in a manner that is satisfactorily analytical, owing primarily to the transcendental mean-field equations (Eqn.~\ref{eqn:Pop_MeanField}) that must be solved numerically. That said, the implication of the sloppy and stiff coordinate observations is that an adjustment of the membrane and synaptic timescales tends to have a larger effect on the large-population fluctuations than it does on the means.

At the end of our isKL analysis, we began a discussion regarding the dimensionality of the models, their behavior, and their manifolds. The largest reductions in the size of the model being discussed occur when moving to population-averaged models and when focusing on the stationary distribution. The combined effect decreases the dimensionality of the behavioral space being studied from $NT/\Delta t$ to $N_{\rm pop}(N_{\rm pop}+3)/2$, in which we essentially shift from a study of particular spike patterns to a study of probability distributions. From here, the isKL methods embed the distribution in $N_{\rm pop}(N_{\rm pop}+3)$ dimensions. This sets the upper limit of dimensionality at the end of our analysis, that upper limit being 18 for the particular architectures studied here. Using an altered participation ratio to measure the effective dimensionality of our embedded spiking models gave us a range of $6\lessapprox PR\lessapprox 8$---depending on the E/I measure $R$---less than the maximum possible dimension. It was illustrated elegantly through the example embedding of the 1-dimensional Poisson model that the participation ratio measures the number of dimensions needed to hold a sufficiently representative version of the model manifold rather than the intrinsic dimensionality of the manifold. In fact, the participation ratio for both toy models indicated that they basically ``filled'' their respective embedding spaces. Taken together, these show that approximated spiking models are definitely undergoing a degree of dimensionality reduction as they are not filling the embedding space like the toy models did. The participation ratio can then be interpreted as giving a sense of how ``pointed'' a change in parameters is. If a 6-dimensional space is needed to represent most features of the manifold, this likely implies that the modulated parameters are mostly affecting 3 natural parameters. However, this is a measure of the effect of base parameter (i.e.~$\tm$ and $\ts$) on behavior, and not necessarily reflect a minimal structure in the base parameter space need for nearly-full expressivity of the model.

From the copious stick-like projections seen in the hierarchies (Figs.~\ref{fig:ForwardHierarchy},\ref{fig:ReverseHierarchy}), we may intuit that the embedded manifolds are of an even smaller dimension than is represented by the participation ratio. We can take this a step further by understanding the entire embedding process as a transformation of a manifold originally in the parameter space, implying that it should intrinsically be, at most, 2-dimensional. We also noted in Sec.~\ref{sec:ParamTransforms} that there are \emph{ad hoc} ways of reducing the parameter space to a 1-dimensional curve while seemingly preserving much of the variability in statistical parameters. If the goal is to find a reduced number of base parameters to approximately cover the manifold in a more principled way, this would likely require estimating the intrinsic dimensionality of the model manifolds with more sophisticated tools than those discussed here. This would provide a number of parameters---or parameter combinations---needed to understand and express the model. Thus, a combination of both an intrinsic measure and the participation ratio provides a complimentary understanding of model manifold dimensionality through the lenses of necessary base parameters and range of impact, respectively. 

Lastly, the properties of these spiking model manifolds is likely to be impacted by the conditions under which it is being studied, more specifically any particular task in which it is being implemented. The models studied here are functionally in a spontaneous regime with a tonic drive that is minor in the scale of the network. The structure of a given task is known collapse high-dimensional spontaneous activity into a lower-dimensional behavioral space \cite{GaoEtAl2017,GanguliEtAl2008}, which might be seen directly in information-geometric interrogations such as the one performed in this paper. Furthermore, this may well affect which statistical parameters are important, in turn changing the coordinate rankings, projection hierarchies, and potentially even the degree to which the resulting manifolds are hierarchical. These possibilities require their own attention in follow-up work.

\subsection*{Limitations}
It is important to discuss the limitations of the framework of modeling and analysis expounded upon in this paper. The primary hurdle to expanded usage of these methods is that the base calculations required for each step combined with the number of samples needed to visually resolve the embedded manifolds make it costly to increase the dimensionality of the parameter space or the number of network elements. The manifolds embedded here required a large number of sampled parameter points to resolve adequately, restricting the number of parameters considered. Similarly, calculating $\sim N^2$ statistical parameters under the Gaussian process approximately would be computationally infeasible and nigh intractable. The first of these restrictions led to the choice of only two key parameters---the timescales---in the current work. The second restriction motivates the reduction of the model by population-averaging. The embedding of the inverse timescale sub-plane (see Sec.~\ref{sec:TimescaleSampling}) revealed that much of the manifold was comprised of points near the boundaries where the behavior became pathological (data not shown). This suggests that a principled or data-informed restriction of parameter space may lead to decrease in the necessary per-parameter sampling density and ease the restrictions presented here.

\subsection*{Future directions}
We conclude by discussing future directions for this work. As developed here, the methods discussed could be applied to models of particular neural circuits in the brain to understand their stationary behavior. For example, one could study the range of behavior of a cortical column when its internal timescales are subject to change. Further, one could study the conditioned range of behaviors in a network in response to a well-defined distribution of inputs as the statistics of the input distribution change. This latter example is meant to demonstrate that the general framework---marginalize, approximate, population-average, and then embed---can apply to modulated parameters other than those presented here.

Perhaps more interesting are the possible extensions of the methods themselves. Of primary interest is the extension of the isKL embedding methods to non-stationary systems. A first-pass way to do this would be to discretize time, apply the embedding procedure at each time-step, and trace points through the embedding space. While conceptually straightforward, this approach would involve significantly more computational investment and the interpretation of the results would be more complicated than in the system discussed in this paper. One could instead try to extend the iSKL embedding framework to apply directly to the path integral representations used to derive the Gaussian process approximations. This would require a proof that the desired properties of the iSKL embedding still hold in this functional context, a potentially harder barrier to clear. Together, these highlight the care with which these conceptual extensions of the current method must be carried out.

\appendix
\begin{widetext}

\section{Population averaging of the Gaussian process approximated network}\label{sec:GPA_PopAvg}
Here, we derive the reduced model from Eqn.~\ref{eqn:PopGPA} by first making a Gaussian process approximation on the full-network spiking model and then averaging the resulting dynamics by population. The stochastically-spiking full network, modeled using a nonlinear Hawkes process, is reproduced here:

\begin{subequations}
\begin{equation*}
    \frac{d V_i}{dt} = -\tm^{-1}(V_i-\varepsilon_I)+I_{i}+\ts^{-1}\left(\mu_{\rm ext}-J_{\rm  self}\dot{n}_i(t)+\sum_{j}w_{ij}\dot{n}_{j}(t)\right)
\end{equation*}
\begin{equation*}
 \dot{n}_i(t)dt\sim {\rm Poiss}[\phi(V_i(t))dt]
\end{equation*}.
\end{subequations}
Recall that the lowercase subscripts ($i$, $j$, etc.) denote individual neurons within the network. $V_i$ is the membrane potential of neuron $i$, $\varepsilon_i$ is the leak reversal potential, $w_{ij}$ is the strength of a synaptic connection from neuron $j$ to neuron $i$, and $-J_{\rm self}$ is an inhibitory self-coupling. The two currents $\mu_{\rm ext}$ and $I_{i}$ represent an average current from an external network and an experimentally injected current, respectively. The process $\dot{n}_i(t)$ is the spike train of neuron $i$, and $\phi(\cdot)dt$ is the instantaneous firing rate nonlinearity, here given by $\phi(x) = \frac{1}{2}(x+\sqrt{x^2+1/2})$. Finally, $\tm$ and $\ts$ are modulated membrane and synaptic timescales, respectively.  The mean-field equations for the steady state membrane potentials can be obtained directly from these equations by using the fact that the approximation neglects fluctuations, and hence $\left \langle n_i(t) \right\rangle =\left \langle \phi(V_i(t)) \right\rangle \approx \phi(\langle V_i(t) \rangle)$, yielding
\begin{equation}
    V_i^{\rm mf}=\varepsilon_I+\tm I_{i}+\frac{\tm}{\ts}\left(\mu_{\rm ext}-J_{\rm self}\phi(V_i^{\rm mf})+\sum_{j}w_{ij}\phi(V_j^{\rm mf})\right).\label{eqn:appendixMF}
\end{equation}
To obtain the dynamics of fluctuations around the mean-field predictions, and to set up for future calculations that go even beyond the Gaussian approximation, it is useful to introduce a path integral representation of this stochastic process, using techniques from statistical physics \cite{ChowBuice2015}. In discrete time, we can write the joint probability for the membrane potential $\mathbf{V}(t)$ and the spike trains $\dot{\mathbf{n}}(t)$ as follows:

\be
P[\mathbf{V}(t),\dot{\mathbf{n}}(t)] = \prod_{t,i}P[V_i(t)|\dot{\mathbf{n}}(t-dt)(t-dt)]P[\dot{n}_i(t-dt)|\mathbf{V}(t-dt)],\nonumber
\ee
where the dynamics of the membrane potential are deterministic given a particular history of the spike trains,
\be
P[V_i(t)|\dot{\mathbf{n}}(t-dt)] \propto \delta\left(\frac{d V_i}{dt}+\tm^{-1}(V_i-\varepsilon_I)-I_{i}-\ts^{-1}\left(\mu_{\rm ext}-J_{\rm  self}\dot{n}_i(t)+\sum_{j}w_{ij}\dot{n}_{j}(t)\right)\right).\nonumber
\ee \normalsize
Here, the proportionality hides a Jacobian factor that arises from a change of variables from $V_I(t)$ to $\dot{V}_I(t)$; this factor is constant for an It\^{o} time discretization, which we assume here.

Next, we take the spike train process to be conditionally Poisson given the current value of the membrane potentials

\be 
P[\dot{n}_i(t-dt)|V_i(t-dt)] = \frac{\phi(V_i(t-dt))^{\dot{n}_i(t-dt)dt}}{(\dot{n}_i(t-dt)dt)!}e^{-\phi(V_i(t-dt))dt},\nonumber
\ee
giving an overall representation

\begin{align*}
P[\mathbf{V}(t),\dot{\mathbf{n}}(t)] = \prod_{t,i}&\delta\left(\frac{d V_i}{dt}+\tm^{-1}(V_i-\varepsilon_I)-I_{i}-\ts^{-1}\left(\mu_{\rm ext}-J_{\rm  self}\dot{n}_i(t)+\sum_{j}w_{ij}\dot{n}_{j}(t)   \right)\right)\nonumber\\
&~~~~~~~~~\times\left[\frac{\phi(V_i(t-dt))^{\dot{n}_i(t-dt)dt}}{(\dot{n}_i(t-dt)dt)!}e^{-\phi(V_i(t-dt))dt} \right].
\end{align*}

In order to cast this in a path integral representation, the standard approach is to represent the probability distributions in terms of a Fourier space representation. For the $\delta$-distribution we have
\be
\delta(x)=\int_{-i\infty}^{i\infty}\frac{d\Tilde{x}}{2\pi}e^{-\Tilde{x}x},\nonumber
\ee
and for a Poisson distribution with rate $\lambda$ we have
\be
p(x) = \int_{-i\infty}^{i\infty}\frac{d\Tilde{x}}{2\pi}e^{-\Tilde{x}x+W(\Tilde{x})}=\int_{-i\infty}^{i\infty}\frac{d\Tilde{x}}{2\pi}e^{-\Tilde{x}x+\lambda(e^{\Tilde{x}}-1)},\nonumber
\ee
where $W(\Tilde{x}) = \lambda(\exp(\tilde{x}) - 1)$ is the cumulant generating function for the Poisson process. We have adopted the standard notation from physics of writing the auxiliary variables this process introduces with tildes, and absorbing the factor of the imaginary unit $i$ into the notation (giving imaginary units of integration). The path integral representation of the spiking process above is then given by
\be
P[\mathbf{V}(t),\dot{\mathbf{n}}(t)] = \int\mathfrak{D}[\Tilde{\mathbf{V}},\Tilde{\mathbf{n}}]e^{-S[\Tilde{\mathbf{V}},\mathbf{V},\Tilde{\mathbf{n}},\dot{\mathbf{n}}]},\nonumber
\label{eqn:pathint_app}
\ee
where $S[\Tilde{\mathbf{V}},\mathbf{V},\Tilde{\mathbf{n}},\dot{\mathbf{n}}]$ is referred to as the ``action'' of the process. We take the continuous-time limit, converting the product over time into an integral over time in the exponent. For this particular model, the action is given by
\small
\begin{align*}
    S[\tilde{\mathbf{V}},\mathbf{V},\tilde{\mathbf{n}},\dot{\mathbf{n}}]=& \int dt\,\,\sum_{i=1}^n\left\{\tilde{V}_i\left [\dot{V}_i+\frac{V_i-\varepsilon_i}{\tm}-I_{i}-\ts^{-1}\left(\mu_{\rm ext}-J_{\rm self}\dot{n}_i(t)+\sum_{j}w_{ij}\dot{n}_{j}(t)  \right) \right]\right.\\
    &~~~~~~~~~~~~~~~~~~~~~~~~~~~~~~~~~~~~+ \tilde{n}_i(t)\dot{n}_i(t)- \left( e^{\tilde{n}_i(t)}-1\right) \phi(V_i)   \Bigg \}.
\end{align*}
\normalsize

For our purposes, it will be convenient to marginalize out the dynamics of the spiking process $\dot{\mathbf{n}}(t)$ and its conjugate variable $\tilde{\mathbf{n}}(t)$ to obtain a representation for the stochastic dynamics of the membrane potentials (along with their auxiliary variables $\tilde{\mathbf{V}}(t)$). The spike-marginalized action is
\begin{align*}
    S[\tilde{\mathbf{V}},\mathbf{V}]=\int dt\,\,\sum_{i=1}^n\left\{\tilde{V}_i\left [\dot{V}_i+\frac{V_i-\varepsilon_I}{\tm} -I_{i}-\ts^{-1}\mu_{\rm ext}\right]-\left(e^{\ts^{-1}\left(-J_{\rm self}\tilde{V}_i + \sum_j\tilde{V}_jw_{ji} \right)}-1\right)\phi(V_i) \right \}.
\end{align*}

The Gaussian process approximation is derived by expanding this action around the mean-field solution, retaining only terms up to quadratic order in $\mathbf{V}(t)-\mathbf{V}^{\rm mf}$ and $\tilde{\mathbf{V}}(t)$. The mean-field solution is obtained by the saddle-points of the action with respect to $\mathbf{V}(t)$ and $\tilde{\mathbf{V}}(t)$, which reproduce Eqn.~(\ref{eqn:appendixMF}) for $\mathbf{V}^{\rm mf}$ and yield $\tilde{\mathbf{V}}^{\rm mf} = \mathbf{0}$. We thus perform a functional Taylor series expansion of the action around  $(\tilde{\mathbf{V}},\mathbf{V})=(\mathbf{0},\mathbf{V}^{\rm mf})$, keeping only terms to the second order in $\delta\mathbf{V}=\mathbf{V}-\mathbf{V}^{\rm mf}$ and $\Tilde{\mathbf{V}}$. The result is
\begin{align*}
S[\tilde{V},V] &= \frac{1}{2}\int dt dt'~ \sum_{ij}\tilde{V}_i(t)\left[  - \ts^{-2}\sum_k \left(-\delta_{ik}J_{\rm self} + w_{ik} \right)\left(-\delta_{jk}J_{\rm self} + w_{jk} \right)\phi(V_k^{\rm mf}) \right]\tilde{V}_j(t') \nonumber\\
&~~~~+\int dt dt'~ \sum_{ij}\tilde{V}_i(t)\left[\delta_{ij}\delta(t-t')\frac{d}{dt} + \delta_{ij}\left(\tm^{-1}  + \ts^{-1}J_{\rm self}\phi'(V_j^{\rm mf})\right) -\ts^{-1}w_{ij}\phi'(V_j^{\rm mf})\right]\delta V_j(t').
\end{align*}
The form of the truncated action is the same as the path integral representation of an Ornstein-Uhlenbeck process derived explicitly by Chow and Buice \cite{ChowBuice2015}. We may therefore match terms to identify the effective stochastic process described by this action:
\be
\frac{d \delta V_i}{dt} = -\sum_{j=1}^n\left[ \delta_{ij}\left(\tm^{-1}  + \ts^{-1}J_{\rm self}\phi'(V_j^{\rm mf})\right) -\ts^{-1}w_{ij}\phi'(V_j^{\rm mf})\right]\delta V_j+\xi_i(t)~~~ {\rm for }~i=1,2,...,n,\nonumber
\ee
where $\xi_i(t)$ is a zero-mean Gaussian noise with covariance
\be
\langle \xi_i(t) \xi_j(t')\rangle = \ts^{-2}\sum_k \left(-\delta_{ik}J_{\rm self} + w_{ik} \right)\left(-\delta_{jk}J_{\rm self} + w_{jk} \right)\phi(V_k^{\rm mf})\delta(t-t').\nonumber
\ee

Casting this as a proper It\^{o} stochastic differential equation, we get
\be
d\delta \mathbf{V} =  -\mathbf{A}\delta \mathbf{V}dt+\mathbf{\Sigma} d\mathbf{W}_t\nonumber
\ee
or equivalently 
\be
d\mathbf{V} =  \mathbf{A}\left(\mathbf{V}^{\rm mf}-\mathbf{V}\right)dt+\mathbf{\Sigma} d\mathbf{W}_t,\nonumber
\ee
where 
\be
A_{ij} = \delta_{ij}\left(\tm^{-1}  + \ts^{-1}J_{\rm self}\phi'(V_j^{\rm mf})\right) -\ts^{-1}w_{ij}\phi'(V_j^{\rm mf})\nonumber
\ee
\be
\left(\mathbf{\Sigma}\mathbf{\Sigma}^T\right)_{ij} = \ts^{-2}\sum_k \left(-\delta_{ik}J_{\rm self} + w_{ik} \right)\left(-\delta_{jk}J_{\rm self} + w_{jk} \right)\phi(V_k^{\rm mf}).\nonumber
\ee
In deriving the reduced dynamics for the population averages, we begin with the Langevin dynamics derived for the full network. We consider the network to have weakly heterogeneous populations in which the connections $w_{ij} = w_{IJ}x_{ij}$ are given by Bernoulli variables, i.e.~$w_{ij}=w_{IJ}x_{ij}$ where $w_{IJ}$ is a constant depending on the pre- and post-synaptic population identities ($J$ and $I$, respectively). We take each connection variable $x_{ij}$ to be independent:
\begin{equation*}
    x_{ij}\sim {\rm Bernoulli}(p_{IJ}).
\end{equation*}
We formally define the average of variable $A_i$ across population $I$ as
\begin{equation*}
    \dlang A_i\drang_I = A_I \equiv \frac{1}{N_I}\sum_{i\in I}A_i(t).
\end{equation*}
At this point, we write the population-averaged connection weights as follows:
\begin{equation*}
    \dlang w_{ij} \drang_I\approx p_{IJ}w_{IJ}.
\end{equation*}
We will derive the effective equations for $V_{I=0} \equiv V_{i=0}$ (the test neuron) and the population averages
\begin{align*}
    V_{I=1} &\equiv \frac{1}{N_1} \sum_{i \in 1} V_i,\\
    V_{I=2} &\equiv \frac{1}{N_2} \sum_{i \in 2} V_i.\\
\end{align*}
We make mean-field-like approximations on the population-average of terms like $\dlang f(A_i) \drang_I\approx f(\dlang A_i\drang_i=f( A_I)$, and we additionally assume approximate independence between the distributions of the synaptic connections, the stationary mean-field potentials $V_i^{\rm mf}$, and the potentials $V_i$. We thus have
\small
\begin{align*}
    \frac{d}{dt}\left(\frac{1}{N_I} \sum_{i \in I} V_i \right) &= \dBlang \sum_j \left[\delta_{ij}\left(\tm^{-1}+\ts^{-1}J_{\rm self}\phi'(V^{\rm mf}_j)\right) - \tau_s^{-1} w_{ij} \phi'(V^{\rm mf}_j)\right] \left(V^{\rm mf}_j - V_j\right) + \xi_i(t)\dBrang_I\\
    &=  \dblang \left(\tm^{-1} +\ts^{-1}J_{\rm self}\phi'(V_i^{\rm mf})\right)V_i^{\rm mf}\dbrang_I -\dblang \left(\tm^{-1} +\ts^{-1}J_{\rm self}\phi'(V_i^{\rm mf})\right)V_i\dbrang_I\\
    &~~~~~~~ -\ts^{-1}\dBlang \sum_j w_{ij}\phi'(V_j^{\rm mf})V_j^{\rm mf} \dBrang_I -\ts^{-1}\dBlang \sum_j w_{ij}\phi'(V_j^{\rm mf})V_j \dBrang_I + \dlang \xi_i(t) \drang\\
    &\approx  \left(\tm^{-1} +\ts^{-1}J_{\rm self}\phi'\left(\dlang V_i^{\rm mf}\drang_I\right)\right)\dlang V_i^{\rm mf}\drang_I  - \left(\tm^{-1} +\ts^{-1}J_{\rm self}\phi'\left(\dlang V_i^{\rm mf}\drang_I\right)\right)\dlang V_i\drang_I    \\
    &~~~~~~~ -\ts^{-1}\dBlang \sum_J N_J\dblang w_{ij}\phi'(V_j^{\rm mf})V_j^{\rm mf}\dbrang_J \dBrang_I -\ts^{-1}\dBlang \sum_J N_J \dblang w_{ij}\phi'(V_j^{\rm mf})V_j \dbrang_J \dBrang_I + \Xi_I(t)\\
    &\approx  \left(\tm^{-1} +\ts^{-1}J_{\rm self}\phi'\left( V_I^{\rm mf}\right)\right) \left( V_I^{\rm mf} - V_I \right)    -\ts^{-1}\dBlang \sum_J N_J\dlang w_{ij}\drang_J\phi'\left(\dlang V_j^{\rm mf}\drang_J\right)\dlang V_j^{\rm mf}\drang_J \dBrang_I\\
    &~~~~~~~~~~  -\ts^{-1}\dBlang \sum_J N_J \dlang w_{ij}\drang_J\phi'\left(\dlang V_j^{\rm mf}\drang_J\right)\dlang V_j \drang_J \dBrang_I + \Xi_I(t)\\
    &\approx  \left(\tm^{-1} +\ts^{-1}J_{\rm self}\phi'\left( V_I^{\rm mf}\right)\right) \left( V_I^{\rm mf} - V_I \right)    -\ts^{-1} \sum_J N_J p_{IJ} w_{IJ}\phi'\left( V_J^{\rm mf}\right)\left( V_J^{\rm mf} - V_J  \right) +\Xi_I(t)\\
    \Rightarrow \frac{dV_{I=1}}{dt} &\approx \sum_{J} \Bigg[\delta_{IJ}(\tau_m^{-1} + \tau_s^{-1} J_{\rm self} \phi'(V_I^{\rm mf})) - \tau_s^{-1} w_{IJ} p_{IJ} N_J \phi'(V^{\rm mf}_J) \Bigg](V^{\rm mf}_J - V_J) + \Xi_I(t). 
\end{align*}
In the last line above, the population-averaged effective noise processes are defined by $\Xi_I(t)=\frac{1}{N_I}\sum_{i\in I}\xi_i(t) $, and the sum over $J$ is over an arbitrary definition of sub-populations. In our particular case, we have $J\in\{0,1,2\}$ as defined in Sec.~\ref{sec:Archs} with $N_0=1$.

Next, we calculate the covariance of the population-averaged noise processes $\Xi_I(t)$. We make the mean-field-like approximations as before:
\begin{align*}
    \left\langle\Xi_I, \Xi_J\right\rangle =& \left\langle\frac{1}{N_I} \sum_{i \in I} \xi_i, \frac{1}{N_J} \sum_{j \in J} \xi_j\right\rangle \\
    & = \frac{1}{N_I N_J} \sum_{i \in I, j \in J}\Big[\langle \xi_i \xi_j \rangle - \langle \xi_i \rangle \langle \xi_j \rangle \Big]\\
    &\approx \frac{\tau_s^{-2}}{N_I N_J} \sum_{i \in I, j \in J}\Bigg[\sum_k \left(-\delta_{ik}J_{\rm self} + w_{ik} \right)\left(-\delta_{jk}J_{\rm self} + w_{jk} \right)\phi(V_k^{\rm mf}) \Bigg]\delta(t-t')\\
    &= \frac{\tau_s^{-2}}{N_I N_J} \sum_{i \in I, j \in J}\Big[\delta_{ij}J_{\rm self}^2\phi(V_{i}^{\rm mf})-J_{\rm self}w_{ji}\phi(V_i^{\rm mf})-w_{ij}J_{\rm self}\phi(V_j^{\rm mf})+ \sum_K\sum_{k\in K} w_{ik} w_{jk}\phi(V_k^{\rm mf})\Big]\delta(t-t')\\
    &=  \frac{\tau_s^{-2}}{N_I N_J} \Big[\delta_{IJ}N_I\dlang J_{\rm self}^2\phi(V_{i}^{\rm mf})\drang_I-N_IN_J\dblang J_{\rm self}w_{ji}\phi(V_i^{\rm mf})\dbrang_{I,J}-N_IN_J\dblang w_{ij}J_{\rm self}\phi(V_j^{\rm mf})\dbrang_{I,J}\\
    &~~~~~~~~~~~~~~~+ N_IN_J \dBlang\sum_K\sum_{k\in K} \dblang w_{ik} w_{jk}\phi(V_k^{\rm mf})\dbrang_K\dBrang_{I,J}\Big]\delta(t-t')  \\
    &= \tau_s^{-2} \Big[\delta_{IJ}\frac{J_{\rm self}^2}{N_I}\phi(V_{I}^{\rm mf})-J_{\rm self}\dlang w_{ji}\drang_{I,J}\phi(V_I^{\rm mf})\\
    &~~~~~~~~~~~~~~~ -\dlang w_{IJ}\drang_{I,J}J_{\rm self}\phi(V_J^{\rm mf})+\sum_K \dlang w_{jk}\drang_{K,I,J}N_K\phi(V_K^{\rm mf})\Big]\delta(t-t')\\
    &= \tau_s^{-2} \Big[\delta_{IJ}\frac{J_{\rm self}^2}{N_I}\phi(V_{I}^{\rm mf})-J_{\rm self}p_{JI}w_{JI}\phi(V_I^{\rm mf}) -p_{JI}w_{IJ}J_{\rm self}\phi(V_j^{\rm mf})+\sum_K p_{JK}w_{JK}N_K\phi(V_K^{\rm mf})\Big]\delta(t-t')\\
    &=\tau_s^{-2}\sum_{K}\left(-\delta_{IK}\frac{J_{\rm self}}{N_K}  +  p_{IK}w_{IK} \right)\left(-\delta_{JK}\frac{J_{\rm self}}{N_K}  + p_{JK} w_{JK} \right) N_K \phi(V^\ast_K)\delta(t-t').
\end{align*}

Note that in this derivation we are assuming an equivalence between the temporal mean-field membrane potential for each individual neuron $V_i$ (used in the previous section) with the mean-field value of the population-averaged membrane potential $V_I$. This amounts to saying the network is sufficiently large and thus the mean of the membrane potential $V_i$ for $i\in I$ tends toward the mean of $V_I$. This yields stochastic differential equation of the form 
\be
d\mathbf{V} =  \mathbf{A}\left(\mathbf{V}^{\rm mf}-\mathbf{V}\right)dt+\mathbf{\Sigma} d\mathbf{W}_t,\nonumber
\ee
where 
\be
A_{ij} = \delta_{ij}\left(\tm^{-1}  + \ts^{-1}J_{\rm self}\phi'(V_j^{\rm mf})\right) -\ts^{-1}w_{ij}\phi'(V_j^{\rm mf})\nonumber
\ee
\be
\left(\mathbf{\Sigma}\mathbf{\Sigma}^T\right)_{ij} = \ts^{-2}\sum_k \left(-\delta_{ik}J_{\rm self} + w_{ik} \right)\left(-\delta_{jk}J_{\rm self} + w_{jk} \right)\phi(V_k^{\rm mf}).\nonumber
\ee

\section{Gaussian process approximation of a population-averaged network}\label{sec:PopAvg_GPA}
In this appendix, we derive the reduced model from Eqn.~\ref{eqn:PopGPA} by first averaging the Hawkes process dynamics across sub-populations and then making a Gaussian approximation, reversing the order of operations in Appendix \ref{sec:GPA_PopAvg}. We begin with the base model:
\begin{subequations}
\begin{equation*}
    \frac{d V_i}{dt} = -\tm^{-1}(V_i-\varepsilon_I)+I_{i}+\ts^{-1}\left(\mu_{\rm ext}-J_{\rm  self}\dot{n}_i(t)+\sum_J\sum_{j\in J}w_{ij}\dot{n}_{j}(t)\right)
\end{equation*}
\begin{equation*}
 \dot{n}_i(t)dt\sim {\rm Poiss}[\phi(V_i(t))dt].
\end{equation*}
\end{subequations}
The population-averaged membrane potential dynamics are given by 
\begin{align*}
    \frac{d}{dt} V_I=&\frac{d}{dt}\dlang V_i\drang_I\\
    & = -\frac{\dlang V_i\drang_I-\varepsilon_I}{\tm}+\dlang I_{i}\drang_I+\frac{\mu_{\rm ext}}{\ts} -\ts^{-1}J_{\rm  self}\dlang\dot{n}_i(t)\drang_I+\ts^{-1}\dBlang\sum_J\sum_{j\in J}w_{ij}\dot{n}_{j}(t) \dBrang\\
    & =-\frac{V_I-\varepsilon_I}{\tm}+I_{I}+\frac{\mu_{\rm ext}}{\ts} -\ts^{-1}J_{\rm  self}\dlang\dot{n}_i(t)\drang_I+\ts^{-1}\sum_J\sum_{j\in J}\dlang w_{ij}\drang \dot{n}_{j}(t).
\end{align*}

As before, we take the connections $w_{ij}$ to be scaled Bernoulli variables, i.e.~$w_{ij}=w_{IJ}x_{ij}$ where $w_{IJ}$ is a constant depending on the pre- and post-synaptic population identities ($J$ and $I$, respectively) and $x_{ij}\sim {\rm Bernoulli}(p_{IJ})$. The population-averaged connections are again given by $\dlang w_{ij}\drang_{I}\approx p_{IJ}w_{IJ}$. We next re-cast the spiking processes into population spiking processes using the following definition
\begin{equation*}
    \dot{m}_I(t)\equiv\sum_{i\in I}\dot{n}_i(t)=N_I\dlang\dot{n}_i\drang_I.
\end{equation*}
As each $\dot{m}_I(t)$ is a sum of conditionally-Poisson processes, it is also a conditionally-Poisson process. Using the same mean-field-esque approximation as before, we may approximate the conditional rate of each $\dot{m}_I(t)$ as follows:
\begin{align*}
    \dot{m}_I=\sum_{i\in I}\dot{n}_i(t)&\sim {\rm Poiss}\left(\sum_{i\in I}\phi(V_i(t))dt\right)= {\rm Poiss}\left(N_I\dlang\phi(V_i(t))\drang_Idt\right)\\
    &\approx {\rm Poiss}\left(N_I\phi(\dlang V_i(t)\drang_I)dt\right)\\
    &= {\rm Poiss}\left(N_I\phi(V_I(t))dt\right).
\end{align*}
With this, the population-averaged Hawkes process dynamics become
\begin{subequations}
\begin{align*}
    \frac{d}{dt} V_I =& -\frac{V_I-\varepsilon_I}{\tm}+I_{I}+\frac{\mu_{\rm ext}}{\ts} -\ts^{-1}J_{\rm  self}\frac{\dot{m}_I(t)}{N_I}+\ts^{-1}\sum_Jp_{IJ}w_{IJ}\dot{m}_J(t)\\
    &~~~~=-\frac{V_I-\varepsilon_I}{\tm}+I_{I} +\ts^{-1}\left(\mu_{\rm ext}+ \sum_J\left(-\delta_{IJ}\frac{J_{\rm self}}{N_I}+p_{IJ}w_{IJ}\right)\dot{m}_J(t)\right)
\end{align*}
\begin{equation*}
 \dot{m}_I(t)dt\sim {\rm Poiss}[N_I\phi(V_I(t))dt].
\end{equation*}
\end{subequations}

After deriving the population-averaged dynamics for the nonlinear Hawkes process, we apply the Gaussian-process approximation scheme to the new dynamics. We begin by applying a mean-field-like approximation to the average of the population-spiking processes, namely $\langle \dot{m}_I(t)\rangle \approx N_I\phi(\langle V_I\rangle)$. This is used to find the stationary mean-field solution for the population-averaged membrane potential dynamics, given by a set of transcendental equations
\begin{equation*}
    V_I^{\rm mf}=\varepsilon_I+\tm I_{I}+\frac{\tm}{\ts}\left(\mu_{\rm ext}+ \sum_J\left(-\delta_{IJ}\frac{J_{\rm self}}{N_I}+p_{IJ}w_{IJ}\right)N_J\phi(V_J^{\rm mf})\right).
\end{equation*}

As in Appendix \ref{sec:GPA_PopAvg}, we represent the joint probability distribution $P[\mathbf{V}(t),\dot{\mathbf{m}}(t)]$ as a path integral by discretizing time, making appropriate Fourier transforms, and taking a continuous-time limit. This yields the expression
\begin{subequations}
    \be
    P[\mathbf{V}(t),\dot{\mathbf{m}}(t)] = \int\mathfrak{D}[\Tilde{\mathbf{V}},\Tilde{\mathbf{m}}]e^{-S[\Tilde{\mathbf{V}},\mathbf{V},\Tilde{\mathbf{m}},\dot{\mathbf{m}}]},\nonumber
    \ee
    with
    \begin{align*}
    S[\tilde{\mathbf{V}},\mathbf{V},\tilde{\mathbf{m}},\dot{\mathbf{m}}]=& \int dt\,\,\sum_{I= 0,1,2}\left\{\tilde{V}_I\left [\dot{V}_I+\frac{V_I-\varepsilon_i}{\tm}-I_{I}-\ts^{-1}\left(\mu_{\rm ext}+ \sum_J\left(-\delta_{IJ}\frac{J_{\rm self}}{N_I}+p_{IJ}w_{IJ}\right)\dot{m}_J(t)   \right) \right]\right.\\
    &~~~~~~~~~~~~~~~~~~~~~~~~~~~~~~~~~~~~+ \tilde{m}_I(t)\dot{m}_I(t)- \left( e^{\tilde{m}_I(t)}-1\right) N_I \phi(V_I)   \Bigg \}.
\end{align*}
\end{subequations}

We marginalize out the explicit spiking dynamics as before by finding the zeros of the derivatives of the action w.r.t.~$\mathbf{\dot{m}}(t)$ and its conjugate variables $\mathbf{\tilde{m}}(t)$. This yields the following marginalized action:
\begin{align*}
    S[\tilde{\mathbf{V}},\mathbf{V}]=&\nonumber\\
    \int dt\,\,&\sum_{I=0,1,2}\left\{\tilde{V}_I\left [\dot{V}_I+\frac{V_I-\varepsilon_I}{\tm} -I_{I}-\ts^{-1}\mu_{\rm ext}\right]-\left(e^{\ts^{-1}\left(-\frac{J_{\rm self}}{N_I}\tilde{V}_I + \sum_J\tilde{V}_jp_{JI}w_{JI} \right)}-1\right)N_I\phi(V_I) \right \}.
\end{align*}

We expand this action around the mean-field solution $(\tilde{\mathbf{V}},\mathbf{V})=(\mathbf{0},\mathbf{V}^{\rm mf})$  to quadratic order. Evaluating individual terms and derivatives at the mean-field solution, we get 
\be
S[\mathbf{0},\mathbf{V}^{\rm mf}]=0,\nonumber
\ee
\be
S_{V_I}[\mathbf{0},\mathbf{V}^{\rm mf}] = 0,\nonumber
\ee
\begin{align*}
S_{\tilde{V}_I}[\mathbf{0},\mathbf{V}^{\rm mf}] =  \int dt\,\,\left[ \dot{V}_i+\frac{V_i^{\rm mf}-\varepsilon_I}{\tm} - \frac{V_i^{\rm mf}-\varepsilon_I}{\tm} \right]=  \int dt\,\,\left[ \dot{V}_i-\dot{V}_i^{\rm mf} \right] =  \int dt\,\,\dot{\delta V}_i, 
\end{align*}
\be
S_{\tilde{V}_I\tilde{V}_J}[\mathbf{0},\mathbf{V}^{\rm mf}] = \int dt\,\, \Bigg[ - \ts^{-2}\sum_K \left(-\delta_{IK}\frac{J_{\rm self}}{N_I} + p_{IK}w_{IK} \right)\left(-\delta_{JK}\frac{J_{\rm self}}{N_J} + p_{JK}w_{JK} \right)N_K\phi(V_K) \Bigg],\nonumber
\ee
\begin{align*}
    S_{\tilde{V}_IV_J}[\mathbf{0},\mathbf{V}^{\rm mf}] = \int dt\,\,\left[ \delta_{IJ}\left(\tm^{-1}  + \ts^{-1}J_{\rm self}\phi'(V_I^{\rm mf})\right) -\ts^{-1}p_{IJ}w_{IJ}N_J\phi'(V_J^{\rm mf})  \right],
\end{align*}
and
\be
S_{V_IV_J}[\mathbf{0},\mathbf{V}^{\rm mf}] = 0.\nonumber
\ee
Again defining fluctuations in the membrane potential as $\delta V_I:=V_I-V_I^{\rm mf}$, approximated action can be written as

\begin{align}
\small S[\tilde{V},V] &= \int dt~\sum_I\left\{ \dot{\delta V}_I + \sum_{J}\left[\delta_{IJ}\left(\tm^{-1}  + \ts^{-1}J_{\rm self}\phi'(V_I^{\rm mf})\right) -\ts^{-1}p_{IJ}w_{IJ}N_J\phi'(V_J^{\rm mf}) \right]\delta V_J \right\}\tilde{V}_I(t) \nonumber\\
&~~~~+ \frac{1}{2}\int dt dt'~ \sum_{ij}\left[ - \ts^{-2}\sum_K \left(-\delta_{IK}\frac{J_{\rm self}}{N_I} + p_{IK}w_{IK} \right)\left(-\delta_{JK}\frac{J_{\rm self}}{N_J} + p_{JK}w_{JK} \right)N_K\phi(V_K) \right]\tilde{V}_I(t)\tilde{V}_J(t') \normalsize.\nonumber
\end{align}
As before, we can identify the GPA dynamics of the population-averaged Hawkes process as corresponding to an Ornstein-Uhlenbeck process. We may therefore match terms to identify the effective stochastic process described by this action:
\be
\frac{d \delta V_I}{dt} = -\sum_{J=0,1,2}\left[ \delta_{IJ}\left(\tm^{-1}  + \ts^{-1}J_{\rm self}\phi'(V_I^{\rm mf})\right) -\ts^{-1}p_{IJ}w_{IJ}N_J\phi'(V_J^{\rm mf})\right]\delta V_J+\xi_I(t)~~~ {\rm for }~I=0,1,2\nonumber
\ee
where $\xi_I(t)$ is a zero-mean Gaussian noise with covariance
\be
\langle \xi_I(t) \xi_J(t')\rangle = \ts^{-2}\sum_K \left(-\delta_{IK}\frac{J_{\rm self}}{N_I} + p_{IK}w_{IK} \right)\left(-\delta_{JK}\frac{J_{\rm self}}{N_J} + p_{JK}w_{JK} \right)N_K\phi(V_K)\delta(t-t').\nonumber
\ee
Casting this as a proper It\^{o} stochastic differential equation, we get 
\be
d\mathbf{V} =  \mathbf{A}\left(\mathbf{V}^{\rm mf}-\mathbf{V}\right)dt+\mathbf{\Sigma} d\mathbf{W}_t,\nonumber
\ee
where 
\be
A_{ij} = \delta_{IJ}\left(\tm^{-1}  + \ts^{-1}J_{\rm self}\phi'(V_I^{\rm mf})\right) -\ts^{-1}p_{IJ}w_{IJ}N_J\phi'(V_J^{\rm mf})\nonumber
\ee
\be
\left(\mathbf{\Sigma}\mathbf{\Sigma}^T\right)_{ij} = \ts^{-2}\sum_K \left(-\delta_{IK}\frac{J_{\rm self}}{N_I} + p_{IK}w_{IK} \right)\left(-\delta_{JK}\frac{J_{\rm self}}{N_J} + p_{JK}w_{JK} \right)N_K\phi(V_K).\nonumber
\ee

We note that this is consistent with the form derived in Appendix \ref{sec:GPA_PopAvg}.

\section{Population averaging for the linear non-spiking model}
We also construct a simpler model of networked, linear non-spiking (or ``graded potential") neurons. We assume the neurons are injected with large numbers of synaptic input that sum together to be approximately Gaussian, with non-zero mean $\mu_{\rm ext}$, creating a stochastic system with dynamics described by
\begin{align*}
     \frac{dV_i}{dt} =& -\tm^{-1}(V_i-\varepsilon_I)+I_i+\ts^{-1}\left(\mu_{\rm ext}-J_{\rm self}\phi(V_i)+ \sum_j w_{ij}\phi(V_j)\right) + \xi_i(t).
\end{align*}
We begin this derivation by assuming the connections $w_{ij}=w_{IJ}x_{ij}$ are scaled Bernoulli variables as in Appendices \ref{sec:GPA_PopAvg},\ref{sec:PopAvg_GPA}. Here, the transfer function $\phi(\cdot)$ is a simple linear function (i.e.~$\phi(x)=x$). The processes $\xi_i(t)$ are zero-mean Gaussian noise synaptic input from neurons external to the network being examined, and thus they scale with $\ts^{-1}$ (i.e.~$\xi_i(t)\sim\ts^{-1}$). We define the covariance of the noise processes $\xi_i(t)$ as follows:
\begin{equation*}
    \langle \xi_i(t)\xi_j(t') \rangle = \ts^{-2}\mu_{\rm ext}\delta(t-t').
\end{equation*}
Here, $k_J$ is a constant potentially depending on the identity of the receiving population $J$. We wish to derive a population-averaged model for the membrane potential dynamics for comparison to the Gaussian-process-approximated spiking models. Again, we define
\begin{equation*}
    \dlang A_i\drang_I \equiv \frac{1}{N_I}\sum_{i\in I}A_i(t).
\end{equation*}
We thus derive the population-averaged dynamics for population $I$:
\begin{align*}
    \frac{d}{dt}\left(\frac{1}{N_I} \sum_{i \in 1} V_i \right) &=  -\dBlang\frac{V_i-\varepsilon_I}{\tm}\dBrang_I+\dlang I_i\drang_I+\ts^{-1}\left(\mu_{\rm ext}-\dlang J_{\rm self}\phi(V_i)\drang_I+ \dblang \sum_J N_J\dlang  w_{ij}\phi(V_j)\drang_J \dbrang_I\right) + \dlang\xi_i(t)\drang_I     \\
    \Rightarrow \frac{dV_{I}}{dt} &\approx -\frac{V_I-\varepsilon_I}{\tm} + I_I  +\ts^{-1}\left(\mu_{\rm ext} - J_{\rm self}\phi(V_I) + \sum_Jp_{IJ}w_{IJ}N_J\phi(V_J)\right) + \Xi_I(t),
\end{align*}
where we have defined $\Xi_I(t) \equiv \frac{1}{N_I} \sum_{i \in I} \xi_i(t)$ for $I=0,1,2$. The means and covariances of the population-averaged noise processes are as follows:
\begin{equation*}
    \langle\Xi_I(t) \rangle = \Bigg \langle\frac{1}{N_I}\sum_{i\in I}\xi_i(t)\Bigg\rangle = \frac{1}{N_I}\sum_{i\in I} \langle\xi_i(t) \rangle = 0,
\end{equation*}
and
\begin{align*}
    \langle\Xi_I(t),\Xi_J(t) \rangle =& \Bigg \langle\frac{1}{N_I}\sum_{i\in I}\xi_i(t),\frac{1}{N_J}\sum_{j\in J}\xi_j(t)\Bigg\rangle = \frac{1}{N_IN_J}\sum_{i\in I,j\in J} \langle\xi_i(t)\xi_j(t) \rangle -\langle\xi_i(t) \rangle\langle\xi_j(t) \rangle\\
    &= \frac{1}{\ts^2 N_IN_J}\sum_{i\in I,j\in J}\delta_{ij}\mu_{\rm ext}\delta(t-t')\\
    &= \frac{\delta_{IJ}}{\ts^2 N_I^2}\sum_{i\in I}\mu_{\rm ext}\delta(t-t')=\frac{\delta_{IJ}}{\ts^2 N_I^2}N_I\mu_{\rm ext}\delta(t-t')\\
    &= \frac{\delta_{IJ}}{\ts^2 N_I}\mu_{\rm ext}\delta(t-t').
\end{align*}
We can then rewrite the population dynamics as
\begin{align*}
    dV_I =& \left(-\frac{V_I-\varepsilon_I}{\tm} + I_I  +\ts^{-1}\left(\mu_{\rm ext} - J_{\rm self}\phi(V_I) + \sum_Jp_{IJ}w_{IJ}N_J\phi(V_J)\right)\right)dt + \Xi_I(t)dt \\
    \rightarrow d\mathbf{V}&=\mathbf{A}\left(\mathbf{A}^{-1}\left(\ts^{-1}\mathbf{\mu}_{\rm ext}+I_{I}\right) - \mathbf{V}\right)dt+\mathbf{\Sigma}d\mathbf{W}_t\nonumber\\
    =&\mathbf{A}\left(\mathbf{\mu} - \mathbf{V}\right)dt+\mathbf{\Sigma}d\mathbf{W}_t,\label{eqn:LFRN}
\end{align*}
where 
\begin{align*}
    \mathbf{A}_{IJ} =& \delta_{IJ}\left(\tm^{-1}-\ts^{-1}J_{\rm self}\right)+\ts^{-1}p_{IJ}w_{IJ}N_J\\
    &= \delta_{IJ}\tm^{-1}+\ts^{-1}w_{IJ}^\ast,\\
    w_{IJ}^\ast &= -\delta_{IJ}J_{\rm self}+ p_{IJ}w_{IJ}N_J,\\
    \left(\Sigma\Sigma^T\right)_{IJ} &= \frac{\delta_{IJ}}{\ts^2 N_I}\mu_{\rm ext}.
\end{align*}

\section{Balance equations}\label{sec:BalEqns}

To derive the balanced state conditions for the network, we begin with the population-averaged spiking network as derived in Appendix \ref{sec:PopAvg_GPA}:

\begin{subequations}
\begin{align*}
    \frac{d}{dt} V_I = -\frac{V_I-\varepsilon_I}{\tm}+I_{I} +\ts^{-1}\left(\mu_{\rm ext}+ \sum_{J=0,1,2}\left(-\delta_{IJ}\frac{J_{\rm self}}{N_I}+p_{IJ}w_{IJ}\right)\dot{m}_J(t)\right)
\end{align*}
\begin{equation*}
 \dot{m}_I(t)dt\sim {\rm Poiss}[N_I\phi(V_I(t))dt],
\end{equation*}
\end{subequations}
where $p_{IJ} w_{IJ}$ came from the population-averaged synaptic connection $\langle\langle w_{ij} \rangle\rangle_J$ and the effective spike count processes are $ \dot{m}_J(t) = \sum_{j\in J}\dot{n}_j(t)$. The total external input to ``neuron'' $I$ is $I_{I} +\tau_s^{-1}\left(\mu_{\rm ext}- \frac{J_{\rm self}}{N_I} \dot{m}_I(t) + \sum_J p_{IJ} w_{IJ} \dot{m}_J(t)\right)$. We want to estimate the mean and variance of this input, taken over the stochastic process. The mean is straightforward, yielding
\begin{align*}
  \tau_s^{-1}\kappa_I &\equiv   I_{I}  + \tau_s^{-1}\left(\mu_{\rm ext}-J_{\rm self}\phi(V_I) + \sum_J p_{IJ} w_{IJ} N_J \phi(V_J) \right).
\end{align*}
Note that the correction term $-J_{\rm self} \dot{m}_I(t)$ is always going to be smaller than the $\sum_J p_{IJ} w_{IJ} \dot{m}_J$ term, so for the purposes of the balanced condition calculation we will neglect it. For the current work, we take the injected currents $I_I$ to be constants.

Calculating the covariance of the total input at times $t$ and $t'$ yields
$$
   \sum_{JK} p_{IJ} w_{IJ} p_{IK} w_{IK} \Big[\langle \dot{m}_J(t) \dot{m}_K(t')\rangle - \langle \dot{m}_J(t)\rangle \langle \dot{m}_K(t')\rangle  \Big].
$$
We make a Poisson approximation to replace the covariance of the $\dot{m}$'s with $\langle \dot{m}_J(t)\rangle \delta_{JK} \delta(t-t')$. Hence, the covariance becomes
$$
    \sum_{J} (p_{IJ} w_{IJ})^2 N_J \phi(V_J) \delta(t-t').
$$
We want the variance of the synaptic input to be $\mathcal O(N^0)$, which means that to leading order we want
$$\sum_J (p_{IJ} w_{IJ})^2 N_J \phi(V_J) \approx (p_{I1} w_{I1})^2 N_1 \phi(V_1) + (p_{I2} w_{I2})^2 N_2 \phi(V_2) \sim \mathcal O(N^0).$$
We neglect the contribution from the test neuron because it is sub-leading here, i.e.~$N_0=1\ll N_1,\,N_2$. In order for this expression to be order $1$, we see that we need $w_{IJ}$ to scale like $1/\sqrt{N}$ as implemented in Eqns.~\ref{eqn:W_NSpk} \& \ref{eqn:W_Spk}.

We return to the mean input to neuron $I$, which we will write as
$$\tau_s^{-1}\kappa_I \approx \sqrt{N}\left( \frac{ I_{I}+\ts^{-1}\mu_{\rm ext}}{\sqrt{N}} + \tau_s^{-1} \left\{ p_{I1}w_{I1} \frac{N_1}{\sqrt{N}} \phi(V_1) + p_{I2}w_{I2} \frac{N_2}{\sqrt{N}} \phi(V_2) \right\} \right).$$
For a balanced network $\kappa_I$ should be $\mathcal O(1)$ for all $I$, which means that the terms in brackets must vanish faster than $1/\sqrt{N}$. We assume that $ I_{I}, \mu_{\rm ext} \propto \sqrt{N}$, and because $w_{IJ} \sim 1/\sqrt{N}$ and $N_I \propto N$ (for $I \neq 0$), the terms in brackets are $\mathcal O(1)$.

As $N \rightarrow \infty$, the terms in brackets must vanish in order for $\kappa_I$ to be finite. This yields a linear system of equations that uniquely determines the means $\mu_I = \phi(V_I)$, and allows us to place constraints on the parameters: 
\begin{equation*}
    -\begin{bmatrix}  I_{1} +\ts^{-1}\mu_{\rm ext} \\ I_{2} +\ts^{-1}\mu_{\rm ext}  \end{bmatrix} = \frac{1}{\ts}\begin{bmatrix} p_{11}w_{11} N_1  & p_{12}w_{12} N_2\\p_{21}w_{21} N_1  & p_{22}w_{22} N_2  \end{bmatrix}\begin{bmatrix} \phi(V_1)\\ \phi(V_2)  \end{bmatrix}.
\end{equation*}

Solving this system of equations for the spike rates $\phi(V_I^{\rm mf})$, we get 
\begin{align*}
    \phi(V_1) &= \frac{\ts}{N_1}\frac{p_{12}w_{12}\left( I_{2} +\ts^{-1}\mu_{\rm ext} \right)-p_{22}w_{22} \left( I_{1} +\ts^{-1}\mu_{\rm ext} \right)}{ p_{11} p_{22} w_{11} w_{22} - p_{12}p_{21}w_{21}w_{12} },\\
    \phi(V_2) &= \frac{\ts}{N_2} \frac{p_{21}w_{21}\left( I_{1} +\ts^{-1}\mu_{\rm ext} \right)-p_{11}w_{11}\left( I_{2} +\ts^{-1}\mu_{\rm ext} \right)      }{ p_{11} p_{22} w_{11} w_{22} - p_{12}p_{21}w_{21}w_{12} }. 
\end{align*}
In the case of our particular models, we can further reduce this expression by noting that $I_I=0$ for $I=1,2$ and $p_{IJ}=p~\forall~I,J$:
\begin{align*}
    \phi(V_1) &= \frac{1}{pN_1}\frac{w_{12} - w_{22} }{w_{11} w_{22} - w_{21}w_{12} }\mu_{\rm ext},\\
    \phi(V_2) &= \frac{1}{pN_2} \frac{w_{21}-w_{11} }{w_{11} w_{22} - w_{21}w_{12} } \mu_{\rm ext}.
\end{align*}
We highlight here that $\phi(V_I)>0$ by its definition as a firing rate. Additionally, $\mu_{\rm ext}$ is assumed to by synaptic input projected into the local network and is thus positive (i.e.~excitatory) here. Taken together, these two points mean the synaptic parameters must satisfy one of the two following sets of inequalities to be in a balanced regime:
\begin{equation}
    \begin{cases}
        w_{11}w_{22} > w_{12}w_{21}\\
        w_{12} > w_{22}\\
        w_{21} > w_{11}
    \end{cases}\label{eqn:balance_conditions1}
\end{equation}
\begin{center}
    or
\end{center} 
\begin{equation}
    \begin{cases}
        w_{11}w_{22} < w_{12}w_{21}\\
        w_{12} < w_{22}\\
        w_{21} < w_{11}
    \end{cases}.\label{eqn:balance_conditions2}
\end{equation}

With this, we have derived the appropriate scaling for the various parameters in the model and found constraints for the synaptic strengths in order satisfy the necessary properties of a balanced network.

\end{widetext}



%

\end{document}